\documentclass[lettersize,journal]{IEEEtran}

\usepackage{amsmath,amsfonts}
\usepackage{algorithmic}
\usepackage{algorithm}
\usepackage{array}
\usepackage{textcomp}
\usepackage{stfloats}
\usepackage{url}
\usepackage{verbatim}
\usepackage{graphicx}
\usepackage{wasysym}
\usepackage{cite}
\usepackage{svg}
\usepackage{booktabs}
\usepackage{multirow}

\usepackage{orcidlink} 
\usepackage{comment}
\usepackage{caption}

\DeclareCaptionFont{smallcaps}{\scshape}

\newcounter{subfig}[figure]
\renewcommand{\thesubfig}{\alph{subfig}}
\newcommand{\subfiglab}[1]{%
  \refstepcounter{subfig}\textbf{(\thesubfig)}\label{#1}%
}

\begin{document}

\title{KinesCeTI: A Modular and Size-Adaptable Force Feedback Glove with Interchangeable Actuation for the Index and Thumb}

\author{Pablo Alvarez Romeo\,\orcidlink{0000-0003-4166-1254} and Mehmet Ercan Altinsoy\,\orcidlink{0000-0002-0803-8818}%
\thanks{P. Alvarez Romeo is with the Centre for Tactile Internet with Human-in-the-Loop (CeTI) and the Chair of Acoustic and Haptic Engineering, Technische Universität Dresden, 01062 Dresden, Germany (e-mail: pablo.alvarez\_romeo@tu-dresden.de).}%
\thanks{M. E. Altinsoy is with the Centre for Tactile Internet with Human-in-the-Loop (CeTI) and the Chair of Acoustic and Haptic Engineering, Technische Universität Dresden, 01062 Dresden, Germany (e-mail: ercan.altinsoy@tu-dresden.de).}%
\thanks{This work involved human subjects or animals in its research. Approval of all ethical and experimental procedures and protocols was granted by the TU Dresden ethic committee under Application No. SR-EK-288072024.}%
}

\maketitle
\begingroup
\renewcommand\thefootnote{}\footnotetext{
This manuscript has been accepted for publication in \emph{IEEE Transactions on Haptics} following peer review. This version corresponds to the accepted manuscript and may differ from the final published version.
}\addtocounter{footnote}{0}
\endgroup
\begin{abstract}
Force feedback gloves in haptic applications remain constrained by limited adaptability, simplified feedback, and fixed architectures that limit force feedback versatility. To address these challenges, we present KinesCeTI, a modular force feedback exoskeleton for the index and thumb, designed as a multipurpose device adaptable to a wide range of hand sizes. The glove incorporates interchangeable thimbles for fingertip or phalanx attachment and a bidirectional tendon transmission that supports both passive and active feedback. It is combined with a modular actuation design, where different feedback systems may be attached. The system was tested with two actuation modules: a compliant ratchet-pawl braking mechanism for passive feedback and a novel one-way clutch for variable active feedback, newly introduced here. The system was evaluated in three user studies with 20 participants each, assessing ergonomics, actuation performance and usability in both real and virtual tasks. Results indicate that the glove adapts to different hand sizes and provides effective feedback with both mechanisms, highlighting its potential as a versatile platform for haptic research.
\end{abstract}


\section{Introduction}

\IEEEPARstart{H}{aptic} technology has received growing attention in recent decades, becoming a key modality in human–machine interaction. However, despite progress \cite{yang2021recent}, realistic haptic feedback remains challenging due to technological limits and the complexity of touch. Applied across diverse domains, haptic interfaces display highly heterogeneous designs \cite{pacchierotti2017wearable,culbertson2018haptics}, with the hands as a primary focus of haptic research due to their central role in manipulation and interaction. Their capabilities have led to the development of a wide range of devices, including thimbles, gloves and externally grounded interfaces, which explore diverse stimulus modalities such as pressure, force, and vibration \cite{pacchierotti2017wearable}. Nevertheless, no standard design has yet emerged.

This work focuses on kinesthetic or force feedback gloves, a class of haptic interfaces widely studied \cite{tiboni2022soft,du2021review,caeiro2021systematic}, although current solutions still face limitations in feedback capabilities, user adaptability and ergonomics. To address these challenges, key design considerations are outlined and a modular, adaptive hand exoskeleton is presented. Its actuation potential is demonstrated through two customized force feedback mechanisms developed in this work and evaluated together with the system’s ergonomics and mobility in three user studies. The results are discussed in the context of the state of the art and the system’s potential as a versatile haptics research platform.

\section{Force Feedback Gloves: Design Considerations and State of the Art}
The design of haptic gloves depends on the features of the human hand. Anthropometric studies report large variability in hand and finger dimensions \cite{chen2011human,greiner1991hand,tilley2001measure}, while the hand is capable of considerable force: mean grip forces up to 55~kg and pinch forces around 10~kg \cite{angst2010prediction}. Regarding mobility, the digits exhibit 21 degrees of freedom (DOFs), with four DOFs per finger and five in the thumb \cite{wang2018toward}. Taking into account such capabilities, the ideal haptic glove would therefore include high-fidelity haptic feedback with low latency, accurate motion tracking below perceptual thresholds, ergonomic design, unobtrusive when inactive, adaptability across hand sizes, ease of use, safety, and affordability. From a design perspective, each desired feature imposes new requirements that are often interrelated and sometimes contradictory. To analyze them, we adopt the architecture described by Wang et al. \cite{wang2018toward}, which divides exoskeletons into five subsystems: sensing, actuation, transmission, control, and structure. Design possibilities for each subsystem are extensive \cite{du2021review,sarac2019design} but interdependent, e.g. the type of control is constrained by sensing and actuation. 

Within this framework, the most restrictive design aspects are motion tracking and
haptics. Regarding feedback modalities, this work focuses exclusively on kinesthetic feedback and does not address tactile or thermal modalities. Even with this limited approach, motion tracking and force feedback must address the multiple DOFs of the hand. In the context of glove design, DOFs can be classified as passive, sensed and feedback DOFs. Each type entails increasing demands in terms of weight, volume, power and complexity. While ideally all DOFs would be sensed and actuated, practical constraints such as limited space, high DOF count, and technological constraints require trade-offs. Moreover, actuated DOFs are not inherently a binary feature (force vs no-force) but involve various qualitative dimensions for complete kinesthetic feedback. Building on prior analyses (Wang et al. \cite{wang2018toward}), this work presents an expanded set of force feedback dimensions:
\begin{itemize}
\item{\textbf{Feedback direction}: some devices apply unidirectional force for a given DOF (e.g., resistance during hand closing), whereas an ideal system enables force in both directions, either alternately or simultaneously.}
\item{\textbf{Transition between states}: seamless and rapid switching between free motion and force feedback, with high backdrivability when no force is required.}
\item{\textbf{Stiffness generation}: rendering stiffness profiles ranging from compliant objects (e.g., foam) to rigid boundaries (e.g., glass surfaces).}
\item{\textbf{Type of force}: passive systems resisting motion (e.g., friction and braking mechanisms) versus active devices capable of inducing movement.}
\end{itemize}

Meeting these dimensions remains challenging, as many design requirements are mutually conflicting. High-fidelity tracking and feedback across all DOFs demand extensive actuation and sensing, yet current technologies lead to heavy, bulky and power-intensive designs that reduce wearability and ergonomics. In general, unidirectional passive systems are simpler, whereas bidirectional active designs are more complex. A further trade-off exists between anatomical adaptability and structural design: soft gloves offer better ergonomics but limited adjustability, whereas rigid exoskeletons can adapt to different hand sizes at the cost of weight, volume and comfort.

Although numerous design strategies have been proposed \cite{pacchierotti2017wearable,du2021review,wang2018toward,iqbal_stroke_nodate}, force feedback gloves inevitably involve trade-offs, exemplified by commercial systems such as HaptX G1 \cite{haptx_glove} and Senseglove Nova~2 \cite{senseglove}. Both adopt soft-glove architectures, providing force feedback through tendon-based transmissions that restrict grasping motion, enabling compact and lightweight designs but limiting force feedback to one direction. SenseGlove and HaptX rely on braking-based actuation, using magnetic friction and pneumatic braking, respectively, to provide passive resistance. Hand-size adaptability is achieved through different glove sizes.

Linkage-based exoskeletons represent an alternative design approach, offering the potential for bidirectional force feedback and cross-user adaptability through an adjustable structure. Dexmo is a prominent example, originally passive \cite{gu2016dexmo} but later extended to active feedback using customized servomotors. SenseGlove’s earlier DK1 also adopted a linkage-based structure, but provided exclusively unidirectional feedback. Beyond fixed-design gloves, Bartalucci et al.~\cite{bartalucci_original_2023} present a bidirectional, hand-size-adaptable glove using a modular architecture with magnetic couplings. However, actuation is offloaded to an external station via Bowden cables, limiting mobility, while the wearable structure weighs approximately 400~g. Chinello et al.~\cite{chinello_modular_2020} also explore modularity through separate kinesthetic and tactile modules. Finger-size adaptability is achieved through a sliding linkage, although kinesthetic feedback is limited to interphalangeal forces. Linkage-based exoskeletons are common in rehabilitation contexts, as exemplified by Iqbal et al.~\cite{iqbal_novel_2014,iqbal_four-fingered_2015}, who implement active bidirectional feedback in direct-drive configurations, and linkage designs adaptable to different hands. In these designs, low backdrivability during free motion is compensated through active control, at the expense of increased power consumption. Conversely, Leonardis et al.\cite{leonardis_hand_2024} present a full upper-limb exoskeleton with haptic thimbles for teleoperation, offering extensive haptic capabilities at the cost of a large, complex, and stationary system with limited portability and wearability. At the opposite end of the design spectrum, some rigid designs prioritize palm-grounded, task-specific feedback, such as Grabity \cite{choi2017grabity}, which simulates weight and opposition grip forces, and Wolverine \cite{choi2016wolverine}, capable of delivering high braking forces with low power consumption.

Across these examples, common limitations of current force feedback gloves emerge. Existing systems tend to fall into two categories: simplified, often unidirectional designs that preserve compactness and ergonomics, or complex and heavy systems that lack adaptability or versatility. Hand-size adaptability is typically addressed through detachable gloves, multiple sizes or, in linkage-based designs, structures with limited adaptability that require prior adjustment or support attachment only at a single point. Moreover, most solutions adopt closed architectures tailored to a single application or feedback profile, reducing versatility and generalizability. Such closed architectures hinder research progress, as haptic researchers often need to develop an entire glove platform before exploring novel concepts. An additional and frequently overlooked factor is acoustic perception: force feedback inherently introduces actuator noise that may influence user perception of both the feedback and the overall interaction and should therefore be considered in the design process.

These limitations highlight the need for a versatile, modular glove architecture that supports multiple force feedback dimensions, adapts to different hand sizes and considers acoustic perception. The following section presents such an exoskeleton, designed with interchangeable actuation and thimble modules and a structure passively adaptable to different users.

\section{KinesCeTI: Design and Actuation Mechanisms}

\subsection{Design goals}
\begin{figure}[!t]
  \centering
  \includegraphics[width=0.8\linewidth]{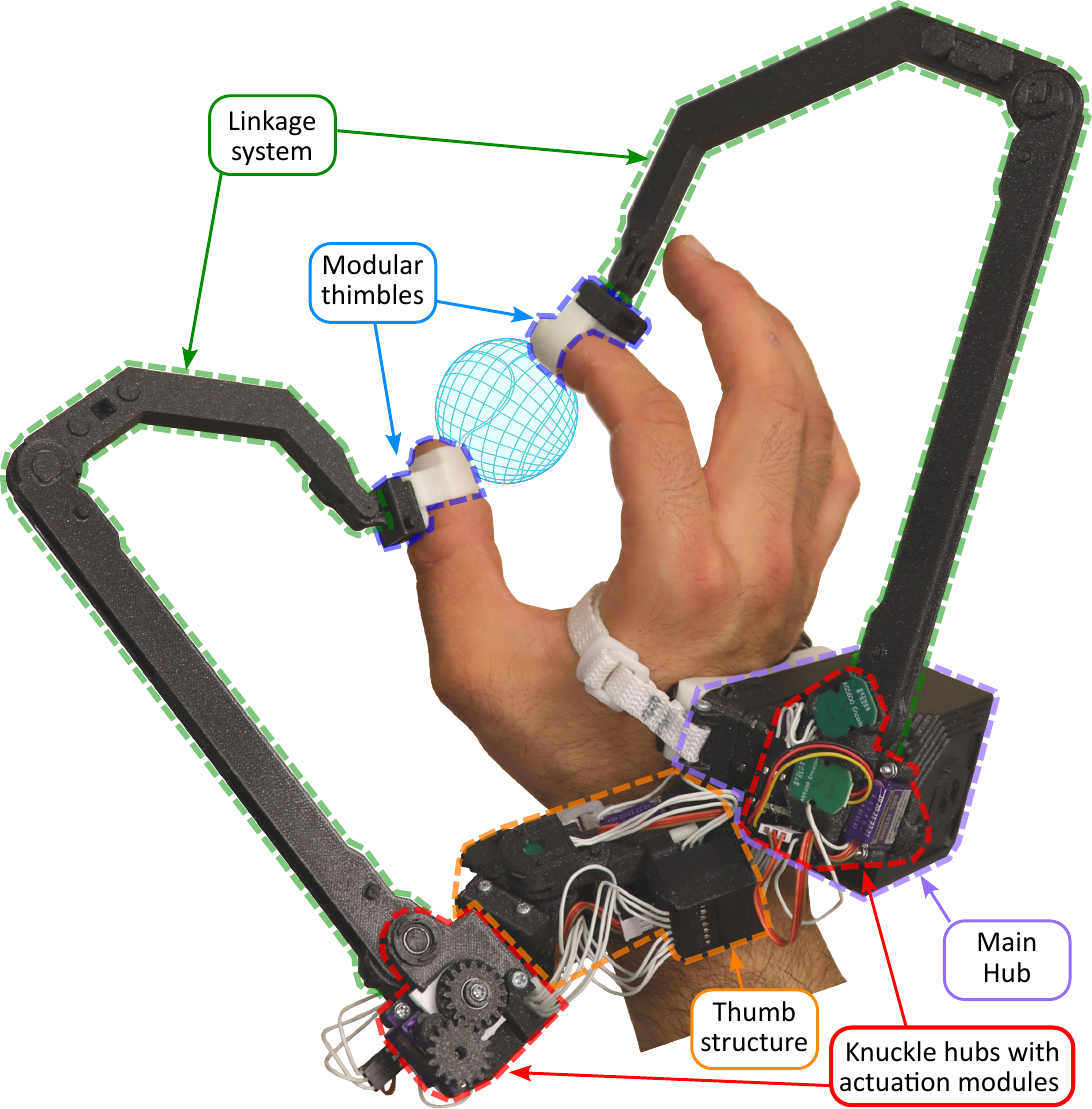}
  \caption{Overview of the KinesCeTI exoskeleton.}
  \label{fig:ExoOverall}
\end{figure}

The previous section reviewed design aspects and highlighted the limitations of current technologies and approaches. Building on this analysis, we present the KinesCeTI exoskeleton (Fig.~\ref{fig:ExoOverall}), designed around the following goals:
\begin{itemize}
\item{Adaptability across adult hand sizes, supporting attachment at either the fingertip or the second phalanx.}
\item{Passively unhindered hand motion with high backdrivability during free motion, ensuring minimal resistance in the absence of force feedback.}
\item{General purpose force feedback device for haptic applications, designed to enable different force feedback dimensions (uni-/bidirectional, active/passive).}
\item{Fully wearable and portable, excluding the need for external actuation stations.}
\item{Low-noise actuation to minimize acoustic annoyance.}
\item{Accessible in terms of materials and components.}
\end{itemize}

\subsection{Overall structure}
Based on these design goals, the choice of subsystems that enable them is addressed, starting with the structural architecture. Although soft actuation systems have advanced significantly in the last decade, they still lack the maturity and accessibility required for bidirectional systems \cite{tiboni2022soft}, as well as hand-size adaptability. A linkage-based architecture mounted on the dorsal side of the hand is therefore selected, enabling hand adaptability while being wearable and portable. As it's grounded to the hand, kinesthetic feedback is limited to finger–hand interactions. Most components are fabricated from common polymer-based materials (PLA, PETG, TPU) to ensure lower costs and accessibility, with the exception of commercial elements such as bearings and electronic components. The exoskeleton is produced via additive manufacturing using Prusa MK3S+ and Prusa Mini+ 3D printers.

To focus on developing an effective design, the exoskeleton targets the index finger and thumb, which together capture the fundamental motion types of the human hand. Since the index is anatomically similar to the middle, ring and little fingers, its design (after further optimizations) can be adapted to those in future iterations. In contrast, the thumb has a unique anatomy and kinematics, requiring a dedicated mechanical design. 

The selected architecture is implemented on a rigid dorsal platform attached to the hand via two adjustable strap bands with buckles positioned around the knuckles and the wrist. For ergonomic purposes, its contact area with the hand consists of a dorsal-shaped TPU structure combined with a neoprene rubber layer for comfortable skin contact. The electronics hub, referred to as the Main Hub, is mounted on this platform. All actuators and sensors are connected to it through different connection boards distributed across the exoskeleton.

\begin{figure}[!t]
  \centering
  \includegraphics[width=1\linewidth]{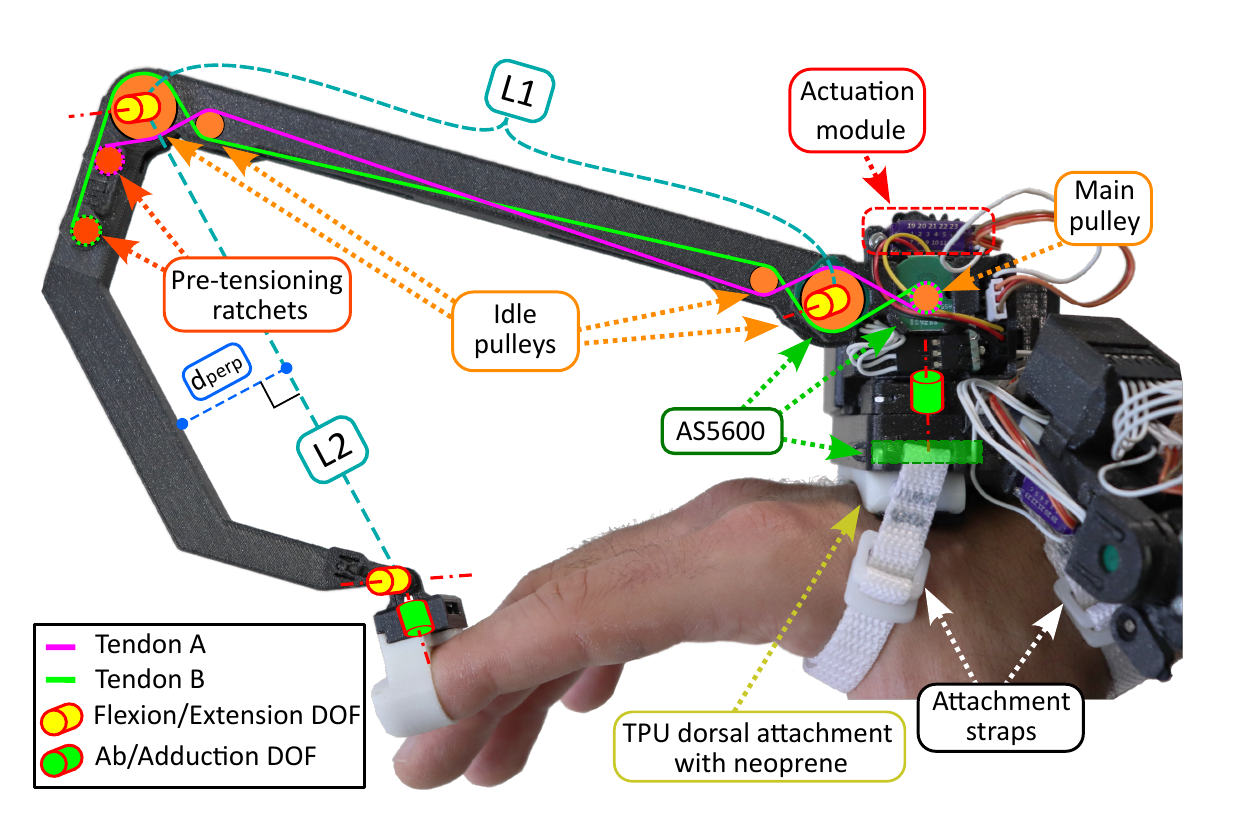}
  \caption{Side view of the index assembly showing the linkage mechanism, force transmission system, and enabled DOFs.}
  \label{fig:TendonRouting}
\end{figure}

The index structure is adjacent to the Main Hub and incorporates a revolute joint that enables ab-/adduction, with its angle measured by a contactless magnetic rotary encoder (AS5600), as shown in Fig.~\ref{fig:TendonRouting}. Mounted above is the Knuckle Hub with the linkage and transmission systems, which are nearly identical for both index and thumb, differing only in linkage lengths, as described in the following subsection. Due to the large motion range of ab-/adduction of the thumb, a structural extension is introduced, oriented at 120$^\circ$ from the index finger in the coronal plane and tilted 30$^\circ$ downward in the sagittal plane (plane conventions follow \cite{bullock2012assessing}).

\begin{figure}[!t]
  \centering
  \includegraphics[width=0.9\linewidth]{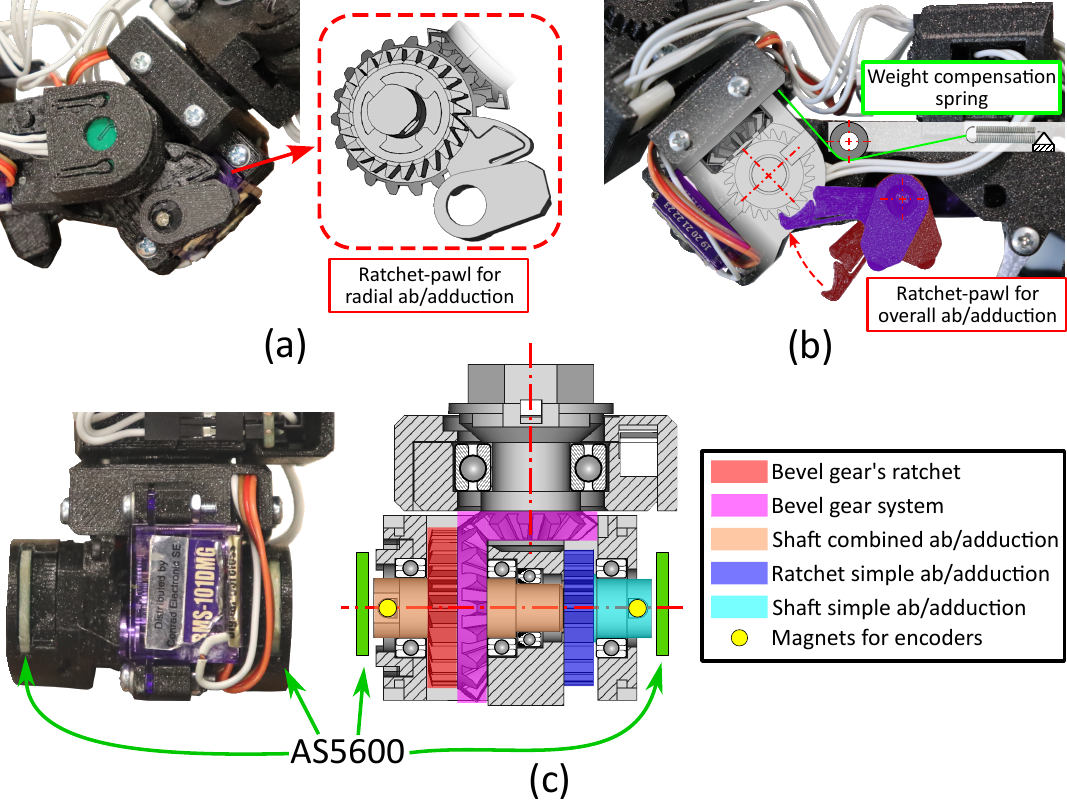}
  \caption{Thumb structure: (a) frontal view of the bevel system with ratchet–pawl mechanism, (b) rear view (housing transparent) of the simple ab-/adduction shaft with ratchet–pawl mechanism, and (c) cross-sectional view showing the main components.}
  \label{fig:ThumbStructure}
\end{figure}

To enable unrestricted thumb motion, the base structure incorporates two sensed and actuated DOFs (Fig.~\ref{fig:ThumbStructure}(c)). The system consists of two concentric shafts, with angular positions measured by two AS5600 encoders mounted on the outer structure. The first DOF corresponds to rotation of the main shaft (simple ab-/adduction), allowing up to $90^\circ$ of motion. The second DOF (combined ab-/adduction) represents the superposition of this main shaft rotation and the ab-/adduction of the Knuckle Hub mounted above it, transmitted through a bevel gear mechanism. The Knuckle Hub itself allows an ab-/adduction range of $\pm90^\circ$ (180$^\circ$ total), with a gear ratio of 13:19 between the Knuckle Hub and the main shaft. 

All rotational systems have mechanical stops at the end of their motion range to ensure safe operation. Both shafts also receive feedback from customized ratchet-pawl mechanisms (Fig.~\ref{fig:ThumbStructure}(a)-(b)), described in the actuation subsection. A compression spring applies a small pulling force to the mobile thumb structure when fully extended, reducing inertia and facilitating user motion.

\subsection{Linkage system}
For both digits, a Knuckle Hub assembly housing the actuators, sensors, and tendon–linkage system is mounted on the ab-/adduction shaft. The two assemblies are mechanically identical, differing only in linkage lengths, whose calculation is explained in this subsection. Correct sizing is required to enable adaptation to different hand sizes. Human hand motion and dimensional variability are well documented in the literature, and multiple references are used \cite{chen2011human,greiner1991hand,ueki2010development,garrett1971adult,ozsoyattempt,ash1996proximal,kong2017investigation,jones2006human}. It should be noted, however, that such datasets do not cover the entire human population, as they are limited to specific countries, age groups, or professions.  

The objective is to identify linkage length combinations that accommodate most human hands across the full range of digit motion and allow attachment either at the fingertip or the second phalanx, enabling a general-purpose glove. The mechanism consists of two serial linkages connected by revolute joints: the base joint at the knuckle hub ($Exo\ base\ joint$), an intermediate linkage connection ($Common\ joint$), and a distal joint at the thimble attachment point ($Thimble\ joint$). The first linkage ($L1$) is modeled as a straight beam, while the second linkage ($L2$) may be straight or arched.

\begin{figure}[!t]
  \centering
  \includegraphics[width=\linewidth]{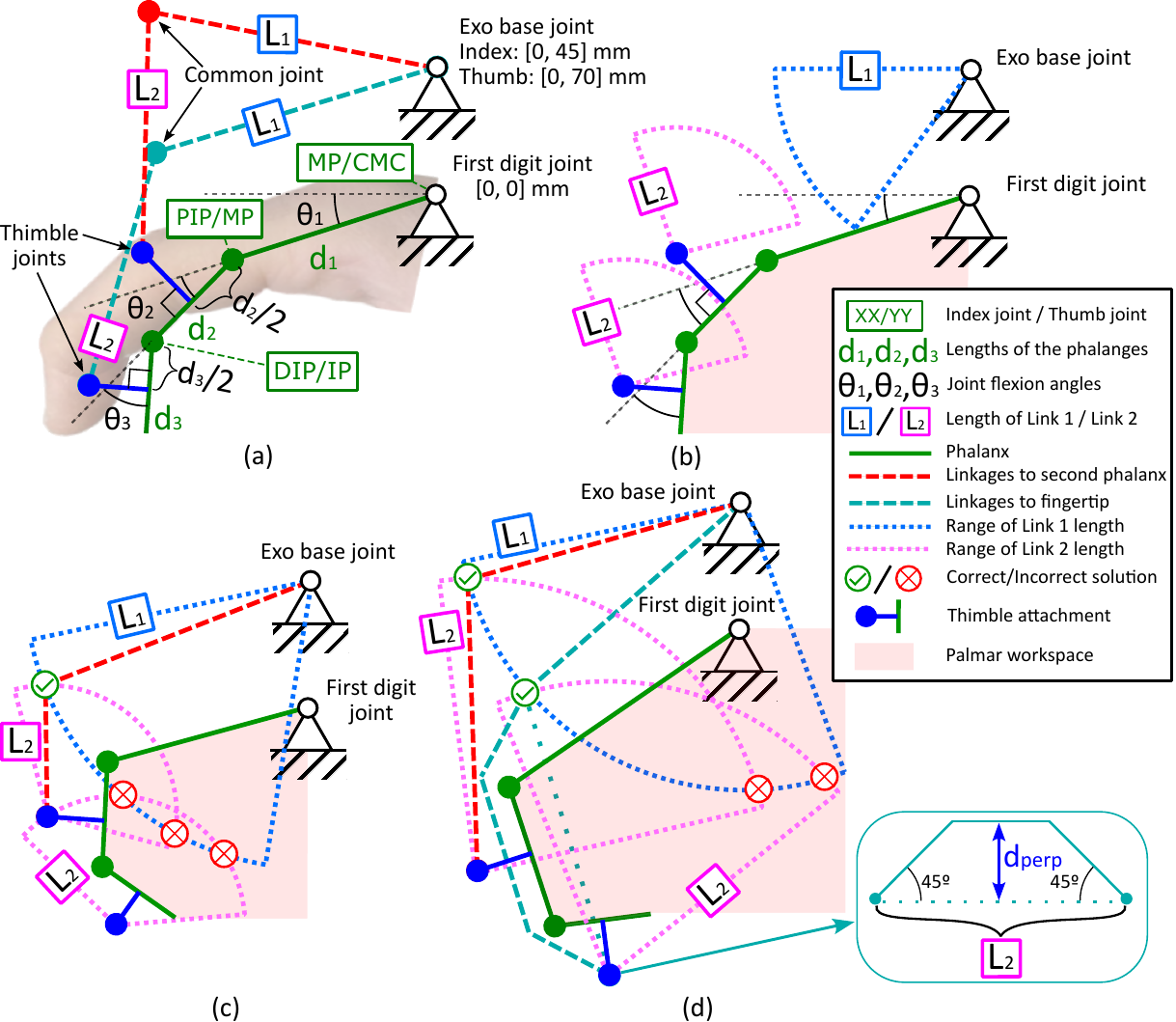}
    \caption{Linkage algorithm: (a) successful configuration with variables indicated, (b) invalid configuration due to insufficient linkage length, (c) invalid configuration caused by intersection with the digit at tip attachment and (d) valid configuration using a modified arched geometry for the second linkage.}
      \label{fig:ExoSizeAlgorithm}
\end{figure}

The analysis considers three representative phalanx size sets, spanning the minimum, mean, and maximum finger dimensions reported in the literature and denoted as $small$, $medium$, and $large$ (Table~\ref{tab:phalanx_lengths}). Digit joint angles are sampled in $10^\circ$ increments over their physiological ranges (Table~\ref{tab:joint_ranges}). The thimble is assumed to be attached at either the fingertip or the second phalanx, with a fixed perpendicular offset of 20~mm (thimble height). Candidate linkage length combinations $(L1, L2)$ are evaluated over 60–200~mm in 10~mm steps. For each linkage length combination and digit configuration, the algorithm first verifies whether both linkages can geometrically form a common joint, and then checks whether the resulting configuration spans the full joint angle range without intersecting the digit geometry. If a collision is detected (Fig.~\ref{fig:ExoSizeAlgorithm}(c)), an alternative configuration with an arched second linkage is evaluated, with arch heights up to 30~mm in 10~mm steps. Linkage combinations that fail to form a valid joint ((Fig.~\ref{fig:ExoSizeAlgorithm}(b))) or to satisfy the motion coverage and collision constraints across all configurations are rejected. 

\begin{table}[!t]
\caption{Phalanx Length Sizes (mm)}
\label{tab:phalanx_lengths}
\centering
\begin{tabular}{@{}llccc@{}}
\toprule
Size  & Finger & Phalanx 1 & Phalanx 2 & Phalanx 3 \\
\midrule
\multirow{2}{*}{Small}  & Index & 35.4  & 14.0  & 10.8  \\
                        & Thumb & 57.0  & 10.0  & 22.0  \\
\midrule
\multirow{2}{*}{Medium} & Index & 58.36 & 21.75 & 26.76 \\
                        & Thumb & 78.57 & 20.02 & 32.40 \\
\midrule
\multirow{2}{*}{Large}  & Index & 100.0 & 32.0  & 36.0  \\
                        & Thumb & 110.0 & 31.0  & 45.0  \\
\bottomrule
\end{tabular}
\end{table}

\begin{table}[!t]
\caption{Digit Joint Motion Ranges and Base Coordinates}
\label{tab:joint_ranges}
\centering
\begin{tabular}{@{}lcccc@{}}
\toprule
Finger & MCP/CMC & PIP/MCP & DIP/IP & Base joint \\
\midrule
Index & $0$--$90^\circ$  & $0$--$110^\circ$ & $0$--$90^\circ$  & $[0,\,45]$ mm \\
Thumb & $0$--$70^\circ$  & $0$--$80^\circ$  & $0$--$90^\circ$  & $[0,\,70]$ mm \\
\bottomrule
\end{tabular}
\end{table}

\begin{figure}[!t]
  \centering
  \includegraphics[width=1.0\linewidth]{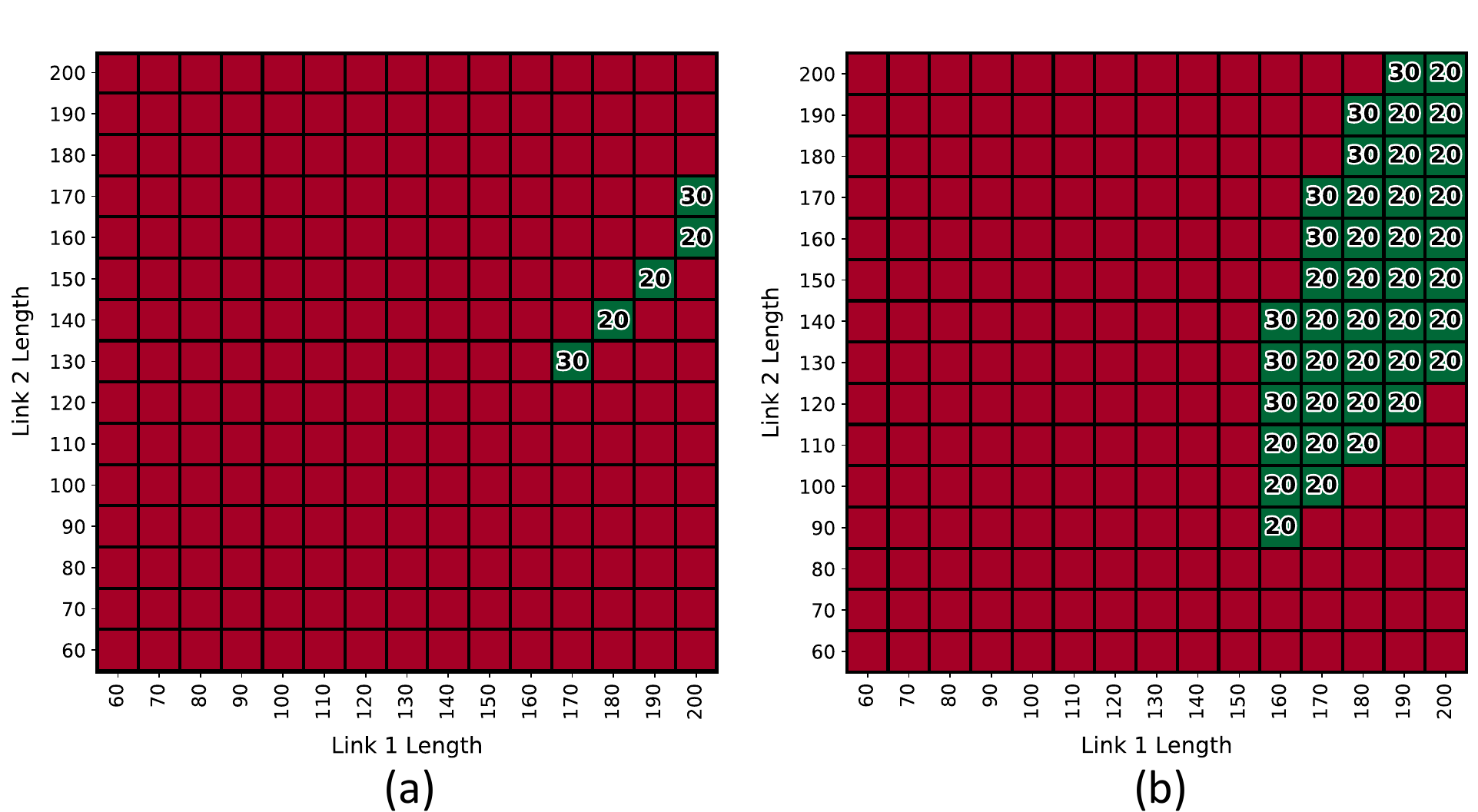}
    \caption{Algorithm results for (a) index and (b) thumb. Green tiles denote successful combinations and their arc height (mm).}
    \label{fig:LengthCombinations}
\end{figure}

From feasible solutions (Fig.~\ref{fig:LengthCombinations}), configurations with the shortest combined length were chosen to minimize weight and volume: L1 = 170~mm, L2 = 130~mm and 30~mm height for the index, and L1 = 160~mm, L2 = 90~mm and 20~mm height for the thumb. To preliminarily assess force capability, quasi-static load tests were conducted using a custom test bench. Each linkage was rigidly clamped at one end, while a perpendicular tensile load was incrementally applied at the distal end via a Spiderwire Dura-4 Braid cable (0.35~mm diameter, 35~kg rated load), connected to a mechanical tuning-gear mechanism and measured using a calibrated scale gauge. The maximum sustained force prior to structural failure was recorded.

For the first linkage, only the index configuration (L1 = 170~mm, corresponding to the longest case) was tested, withstanding an ultimate tensile load of 52.6~N. For the second linkage, the index configuration (L2 = 130~mm) withstood 45.3~N, while the thumb configuration (L2 = 90~mm) sustained 62.5~N. These tests represent preliminary feasibility checks rather than definitive strength measurements.

\subsection{Modular thimbles}
\begin{figure}[!t]
  \centering
  \includegraphics[width=0.9\linewidth]{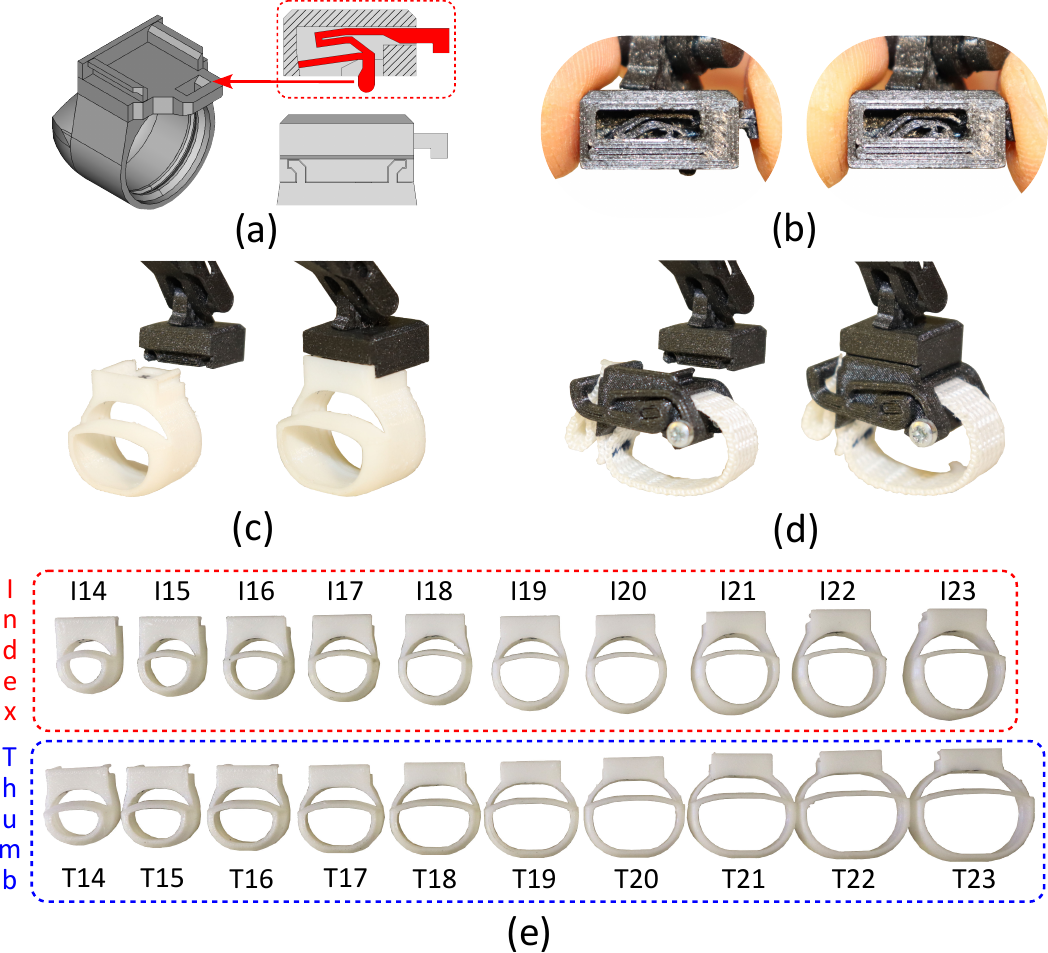}
  \caption{Thimble system: (a) compliant latch design and placement, (b) latch operation, (c) TPU fingertip thimble and attachment, (d) strap thimble and attachment, and (e) TPU fingertip thimble designs for index and thumb.}
  \label{fig:Thimble}
\end{figure}

The second linkage terminates in a thimble assembly that serves as the end effector and provides two revolute joints for flexion/extension and ab-/adduction. Although the latter is already present at the knuckle joint, this additional DOF at the thimble ensures alignment with the digit axis, compensating for plane misalignment between digit motion and linkage orientation. The thimble assembly incorporates a slider–lock attachment mechanism comprising an insertion slot and a compliant PETG latch that secures it in place (Fig.~\ref{fig:Thimble}(a)–(b)).

Two thimble designs were developed. The first is intended for fingertip attachment and requires customized geometries, as the fingertip is a free end without an anatomical joint for fixation and exhibits higher geometric variability (Fig.~\ref{fig:Thimble}(c)). For flexibility, the chosen material is TPU 95A, being produced in ten sizes for each digit with inner diameters from 14 to 23~mm (1~mm step). Additional width variations (2–6.5~mm) are implemented for the thumb (Fig.~\ref{fig:Thimble}(e)). The second design is a single-size module for the second phalanx, consisting of a polypropylene strap secured by a PETG hub with an integrated spring latch, which releases when pressed (Fig.~\ref{fig:Thimble}(d)).
  
To evaluate the force capabilities of the thimble assembly, preliminary load tests were performed on both designs using the same test bench described in the previous subsection. For the strap-based design, five samples were tested, withstanding loads of $98.98\pm 14.55$~N before failure due to strap loosening while remaining operational. For the fingertip thimbles ten samples were tested, sustaining $45.42\pm 6.11$~N. In this case, failure occurred when the TPU slider disengaged from the PETG holder, after which the thimble remained functional.

\subsection{Force transmission system}

The linkages, thimbles and base structure form a continuous mechanical chain connecting the digits to the dorsal side of the hand. As this structure is mechanically passive, additional components are required for sensing and force feedback. To minimize weight and inertia at the digits, both actuation and sensing elements are placed at the Knuckle Hubs. Motion and force are transmitted through a bidirectional closed-loop tendon-pulley architecture, also known as a cable-routed pulley (CPR) configuration with endless cables \cite{grosu2018driving}. This architectural choice enables the implementation of the feedback dimensions from Section~2 (passive/active, uni-/bidirectional, variable impedance). Two braided polyethylene tendons (Spiderwire Dura-4 Braid) are used, each anchored at the main pulley on the Knuckle Hub and at a ratchet–pawl pretensioner on the second linkage, where pretensioning is applied.

The tendon routing transmits the rotation of both linkages to the main pulley, while intermediate pulleys guide the tendons within the structure and prevent slack, as illustrated in Fig.~\ref{fig:TendonRouting}. In the absence of force feedback, linkages $L1$ and $L2$ are kinematically independent and can move freely. When force feedback is engaged, the tendon–pulley system kinematically couples their flexion/extension motion. For sensing, two AS5600 encoders are used: one at the first exoskeleton joint to measure the angle of the first linkage ($L1$), and one at the main pulley to capture the combined angle of both linkages, providing two sensed DOFs. 

\subsection{Actuation system}
While the described subsystems allow all force feedback dimensions, the actuation system ultimately determines which can be realized. To this end, the Knuckle Hub features a modular actuation architecture that supports motor systems with varying capabilities. The main pulley includes a socket for attaching external actuators or sensing modules (Fig.~\ref{fig:ActuatorModule}(a)). In this work, two actuation systems were developed and implemented: a ratchet-pawl mechanism and a novel one-way clutch (Fig.~\ref{fig:ActuatorModule}(c)–(e)), both driven by servomotors, specifically the Bluebird BMS-101DMG, selected based on prior acoustic annoyance studies for wearable actuators \cite{alvarez2022comparative,alvarez2023preliminary}. 

Acoustic characterization of the servomotor indicated A-weighted sound pressure levels of 32.3~dB(A) and 44.1~dB(A) during slow and fast unloaded motions ($120^\circ$ in $\approx3$~s and $\approx0.18$~s, respectively), increasing to 54.9--57.3~dB(A) under applied load torques of 0.1--0.4~kg$\cdot$cm \cite{alvarez2022comparative}. Perceived annoyance (0--100 scale) was rated as low to medium in passive scenarios (e.g., $M=40.16$, $SD=23.05$ at 0.1~kg$\cdot$cm) and lower when integrated into a VR setup ($M=26.00$, $SD=23.59$) \cite{alvarez2023preliminary}. These measurements characterize the actuator itself, while overall device acoustics depend on mounting conditions. In addition, the motor is lightweight (4.4~g), provides sufficient torque and speed for the intended application (1~kg$\cdot$cm and 0.07~s/$60^\circ$ at 6~V), and features compact dimensions (18.6~$\times$~7.6~$\times$~15.7~mm). The servomotors were modified to include a connector to the internal potentiometer, enabling internal position sensing for control purposes.

\subsubsection{Ratchet-pawl mechanism}
\begin{figure}[htbp]
  \centering
  \includegraphics[width=0.8\linewidth]{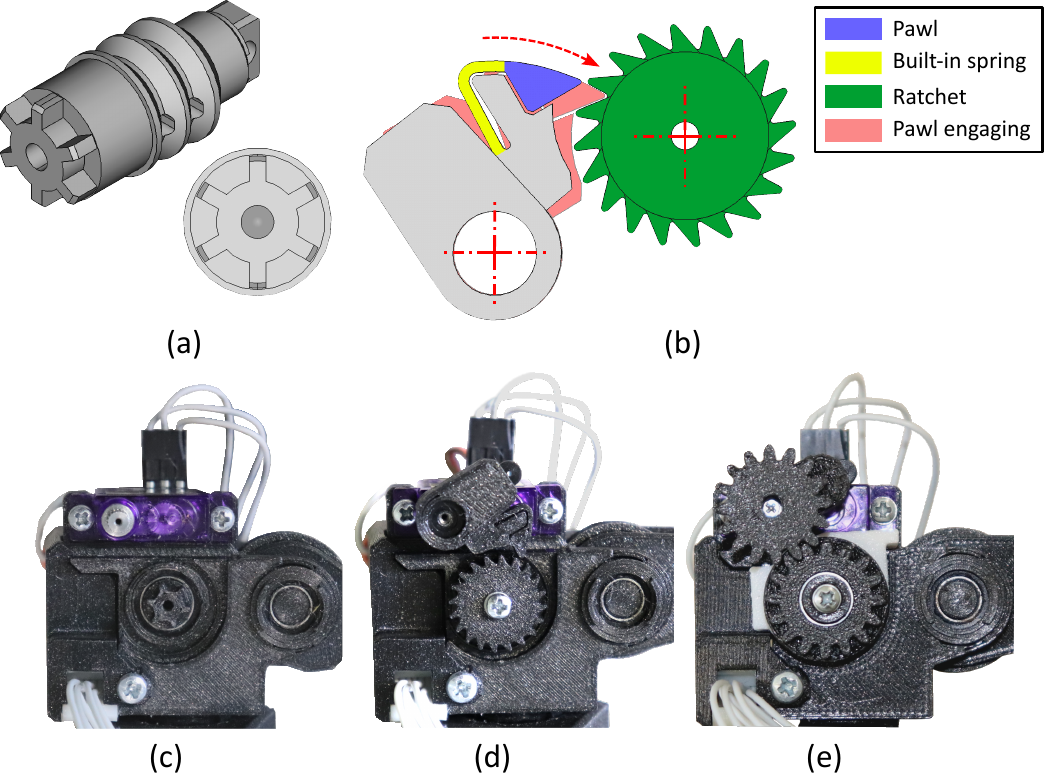}
\caption{Modular actuation system: (a) main pulley socket, (b) ratchet–pawl mechanism, (c) main pulley without mechanism, (d) installed ratchet–pawl, and (e) installed one-way clutch.}
  \label{fig:ActuatorModule}
\end{figure}

The first actuation system is a custom compliant ratchet-pawl mechanism (Fig.~\ref{fig:ActuatorModule}(b)). The ratchet is fixed to the main pulley, while the pawl with a built-in spring is attached to the servomotor, providing binary feedback, allowing the pulley to rotate freely or be blocked. This enables unidirectional stiff boundary force feedback with high backdrivability. Stiffness is inherently high due to its ratchet-pawl principle, but slightly reduced by the overall structure compliance. A trade-off exists between angular resolution, robustness, and clutching speed, as increasing the number of teeth improves resolution but reduces structural strength and is constrained by additive manufacturing limits. With an outer diameter of 16~mm and an inner diameter of 12–13~mm (radial tooth height 1.5–2~mm, face width 4 mm), a 20-tooth configuration ($18^\circ$/tooth) was selected from tested designs (12–30 teeth) as the best compromise. Preliminary load tests on three samples yielded failure loads of 155~N, 157.1~N and 131.8~N, consistently above the maximum thimble loads.

The mechanism is lightweight (pawl, ratchet, and screw: 1.74~g) and simple, also adapted for thumb ab-/adduction feedback (Fig.~\ref{fig:ThumbStructure}(a)-(b)). Clutching latency was characterized with 20 samples per motor, yielding means of $83.15 \pm 33.13$~ms (28–150~ms) for the combined 40 Knuckle Hub samples, $71.65 \pm 29.07$~ms (37–144~ms) for the fixed thumb motor, and $79.15 \pm 20.28$~ms (50–125~ms) for the bevel motor.

\subsubsection{One-way clutch mechanism}
\begin{figure}[t]
  \centering
  \includegraphics[width=0.9\linewidth]{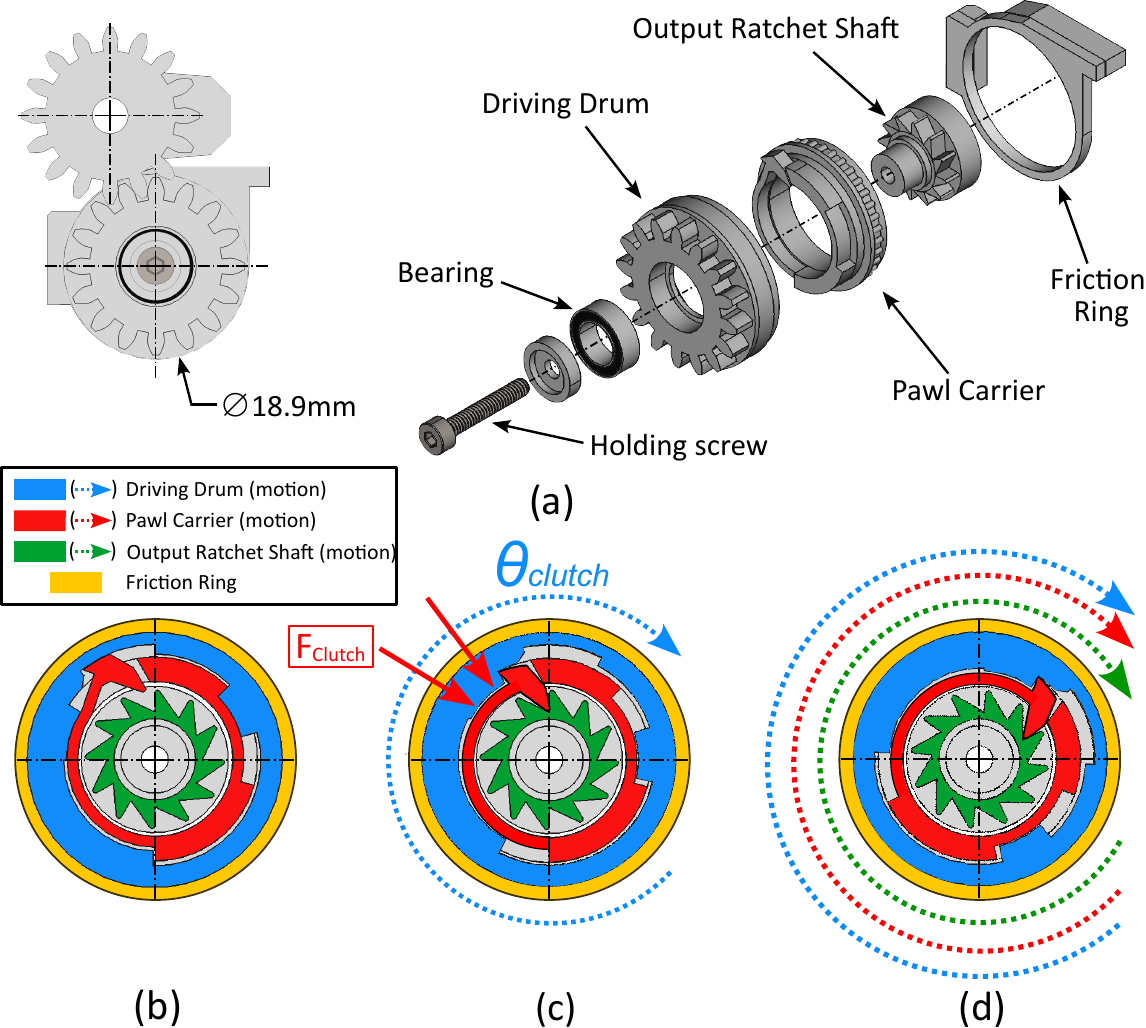}
  \caption{One-way clutch mechanism: (a) frontal and exploded views, (b) Free motion status, (c) Clutching operation, and (d) Feedback mode.}
  \label{fig:OneWayClutch}
\end{figure}

To enable active unidirectional feedback with variable impedance while maintaining good backdrivability, a novel clutch mechanism is developed. In this design, both clutching and feedback are driven by the same servomotor, using an architecture similar to \cite{romeo2021passively} but based on a compliant ratchet–pawl principle. The components, shown in Fig.~\ref{fig:OneWayClutch}(a), are as follows:

\begin{itemize}
\item{\textbf{Output Ratchet Shaft (ORS)}: fixed to the main pulley socket, includes a ratchet.}
\item{\textbf{Pawl Carrier (PC)}: surrounds the ORS and houses a compliant pawl with an increasing radial profile.}
\item{\textbf{Driving Drum (DD)}: encloses both the PC and ORS and is concentrically mounted via a bearing and a screw. It is driven by the servomotor through a direct drive gear.}
\item{\textbf{Friction Ring (FR)}: provides low friction between the PC and the outer structure (Knuckle Hub).}
\end{itemize}

All parts are 3D-printed in PETG, except the FR in TPU. The operation of the clutch, depicted in Fig.~\ref{fig:OneWayClutch}(b), is as follows:

\begin{itemize}
\item{\textbf{Free motion}: ORS and main pulley rotate freely, decoupled from the motor.}
\item{\textbf{Clutching}: when feedback is required, the motor drives the DD while the FR holds the PC. During relative rotation, the fixed inner radius of the DD interacts with the pawl’s increasing radial profile, inducing a radial inward displacement that engages the ratchet.}
\item{\textbf{Feedback}: once engaged, the system enables active feedback, capable of applying resistance or active forces against the digit.}
\item{\textbf{Unclutching}: reverse motor rotation or opposite digit motion disengages the pawl, restoring free motion.}
\end{itemize}

The clutching angle ($\theta_\text{clutch}$) is defined as:
\begin{equation}
\theta_\text{clutch} = \theta_\text{pawl} + \theta_\text{align}
\end{equation}

Here, $\theta_\text{pawl}$ is the rotation required for the DD to push the pawl inward, while $\theta_\text{align}$ is the additional angle needed to align with the ratchet teeth. $\theta_\text{align}$ ranges from $0^\circ$ (perfect alignment) to one ratchet step. In our current design, $\theta_\text{pawl}$ is approximately $30^\circ$ and since the ratchet has 12 teeth, the ratchet step angle is $30^\circ$. Thus, the total clutching angle can theoretically range from $30^\circ$ to $60^\circ$. Clutching latency also depends on finger motion: when the finger is moving, $\theta_\text{align}$ decreases, thereby reducing the time required to engage.  

The clutch mechanism was implemented in the exoskeleton and measured in both the index and thumb Knuckle Hubs, with 20 samples collected in each. The exoskeleton was kept still during the measurements and was not attached to any hand, to avoid finger movements that may affect the latency. Times were measured between the sent command and motion detection using the corresponding AS5600 sensors. For the combined data, mean latency was $100.1 \pm 27.78$~ms (range: 27--160~ms) and mean clutching angles were $62.66 \pm 15.2^{\circ}$ (range: 29.46--97.44$^{\circ}$). The complete clutch module, including bearing and bolt, weighs approximately 4.39~g and measures $\diameter 18.6 \times 14$~mm (excluding the servomotor).
\begin{figure*}[b]
\centering
\begin{minipage}[b]{0.52\textwidth}
\vspace{0pt}
\captionof{table}{Hand anthropometric measurements}
\label{tab:test1_hand_anatomy}
\scriptsize
\setlength{\tabcolsep}{2pt}
\renewcommand{\arraystretch}{1.0}
\begin{tabular}{@{}r p{0.22\textwidth} r r r r p{0.6em} r p{0.22\textwidth} r r r r@{}}
\toprule
\textbf{ID} & \textbf{Description} & \textbf{Mean} & \textbf{SD} & \textbf{Min} & \textbf{Max} &
& \textbf{ID} & \textbf{Description} & \textbf{Mean} & \textbf{SD} & \textbf{Min} & \textbf{Max} \\
\midrule
1  & I-DP length              & 25.04 & 1.81 & 21.3  & 28.58 & & 15 & I-DP circumference             & 50.5  & 3.51 & 44    & 57 \\
2  & I-MP length              & 25.24 & 2.80 & 20.98 & 32.53 & & 16 & I-DIP circumference            & 54.8  & 2.80 & 49    & 60 \\
3  & I-PP length              & 43.66 & 3.84 & 36.47 & 51.61 & & 17 & I-MP circumference             & 59.05 & 2.80 & 54    & 64 \\
4  & I-MC length              & 73.38 & 3.69 & 66.37 & 79.66 & & 18 & I-PIP circumference            & 66.5  & 2.64 & 62    & 71 \\
5  & I PIP-crease length  & 25.61 & 2.35 & 20.29 & 30.15 & & 19 & I-PP circumference             & 69.5  & 3.15 & 62    & 74 \\
6  & MI-MC length              & 67.11 & 4.66 & 59.1  & 75.38 & & 20 & T-DP circumference             & 61.45 & 3.85 & 55    & 69 \\
7  & RI-MC length              & 63.07 & 3.35 & 57.59 & 69.13 & & 21 & T-IP circumference             & 67.25 & 3.52 & 61    & 73 \\
8  & LI-MC length              & 63.05 & 5.43 & 54.73 & 76.62 & & 22 & T-PP circumference             & 69.6  & 3.29 & 62    & 75 \\
9  & T-DP length              & 31.36 & 2.44 & 26.6  & 35.45 & & 23 & Knuckles perimeter  & 205.1 & 11.09& 174   & 221 \\
10 & T-PP length              & 34.91 & 4.29 & 28.08 & 42.3  & & 24 & Palm perimeter      & 218.65& 12.30& 189   & 233 \\
11 & T-MC length              & 50.23 & 4.50 & 42.81 & 59.54 & & 25 & Wrist perimeter     & 174.25& 8.73 & 159   & 188 \\
12 & Hand breadth (MCP)   & 83.89 & 3.94 & 74.32 & 91.15 & & 26 & I length to knuckle*           & 93.93 & 7.32 & 80.31 & 107.61 \\
13 & Hand breadth (palm)       & 86.52 & 4.06 & 76.97 & 91.87 & & 27 & I length to crease*            & 75.88 & 5.69 & 64.34 & 82.79 \\
14 & Wrist breadth      & 60.25 & 3.77 & 52.01 & 69.72 & & 28 & T length to wrist*             & 116.5 & 8.06 & 101.88& 130.07 \\
\bottomrule
\end{tabular}
\vspace{2pt}
\footnotesize\par
Units: mm. \textit{Abbreviations:} I=index, T=thumb, MI=middle, RI=ring, LI=little, DP=distal phalanx, MP=middle phalanx, PP=proximal phalanx, MC=metacarpal, MCP/PIP/DIP/IP=standard joints. *Derived segments, calculated as in Fig.~\ref{fig:test1_visual_anatomy}.
\end{minipage}
\hfill
\begin{minipage}[b]{0.37\textwidth}
\includegraphics[width=\linewidth]{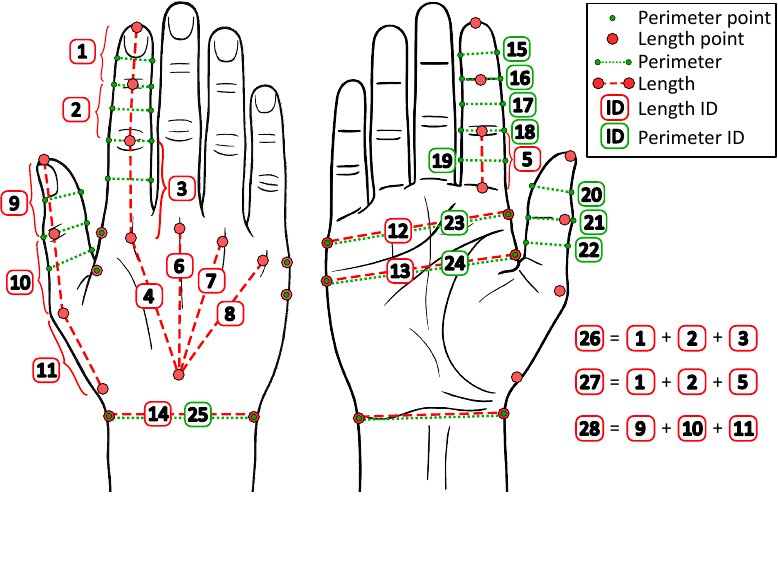}
\captionof{figure}{Dimensions of the hand, according to Table~\ref{tab:test1_hand_anatomy}.}
\label{fig:test1_visual_anatomy}
\end{minipage}
\end{figure*}

\subsection{Control system}
The control system is implemented on an ESP32 DEVKITV1 embedded in the Main Hub, which drives the servomotors and acquires sensor data. Position sensing is provided by seven AS5600 encoders and buffered servomotor potentiometers, while hand orientation is tracked using a 9-DOF IMU (Adafruit BNO055). The system does not include force sensors, and all control algorithms rely exclusively on position measurements from the encoders and servomotor potentiometers. Electronics are powered via USB from the host computer, whereas the four servomotors are powered separately at 6 VDC (up to 1 A) using a Voltcraft SNG-1000-OC. During experiments, the ESP32 communicates over serial with a CHAI3D simulation \cite{conti2003chai} running on the host computer, which performs digit calibration, computes interactions with virtual objects, and provides position-based feedback commands. Further details are presented in the next section.

\subsection{Overall summary}
This section presented the design of a modular and adaptive exoskeleton with a total mass of 376 g, measured with TPU thimbles and servomotors but excluding actuation mechanisms. The system integrates interchangeable actuation modules, modular thimbles and an adaptive linkage structure to fit different hand sizes. Its sensing and actuation architecture enables both binary and variable force feedback, supporting uni- and bidirectional interaction. These features are evaluated in the user studies described in the next section. To facilitate reproducibility, the CAD models, bill of materials, and assembly documentation of the proposed system are publicly available at \href{https://doi.org/10.5281/zenodo.19219658}{https://doi.org/10.5281/zenodo.19219658}.

\section{User studies}
The exoskeleton and its subsystems are evaluated in three user studies:
\begin{itemize}
\item{An ergonomics and range of motion (RoM) study.}
\item{A pick-and-place manipulation task in VR.}
\item{A pinching softness discrimination task in VR.}
\end{itemize}
All studies were conducted with written informed consent and approved by the TU Dresden Ethics Committee (SR-EK-288072024). As the device is designed for right-hand use only, all participants performed the tasks with the right hand, regardless of handedness. For the two VR experiments, a brief automatic software calibration was performed at the beginning, during which participants executed three pinch motions to define the motion range for the virtual digits.

\subsection{Ergonomics and Range of Motion (RoM) study}
\subsubsection{Experiment definition}
\begin{figure*}[t]
  \centering
  \begin{minipage}[t]{0.54\textwidth}
    \centering
    \includegraphics[width=\linewidth]{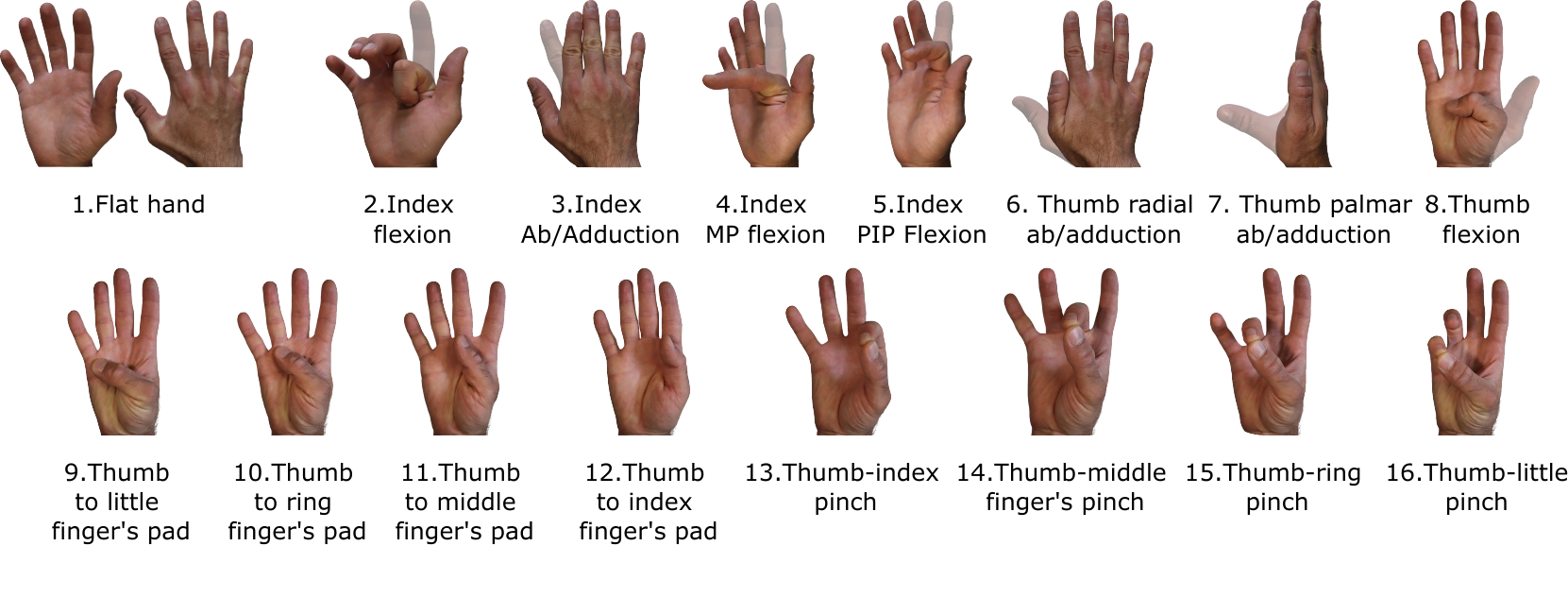}

    \vspace{0.3em}
    \subfiglab{fig:HandPostures}\;
  \end{minipage}\hfill
  \begin{minipage}[t]{0.43\textwidth}
    \centering
    \includegraphics[width=\linewidth]{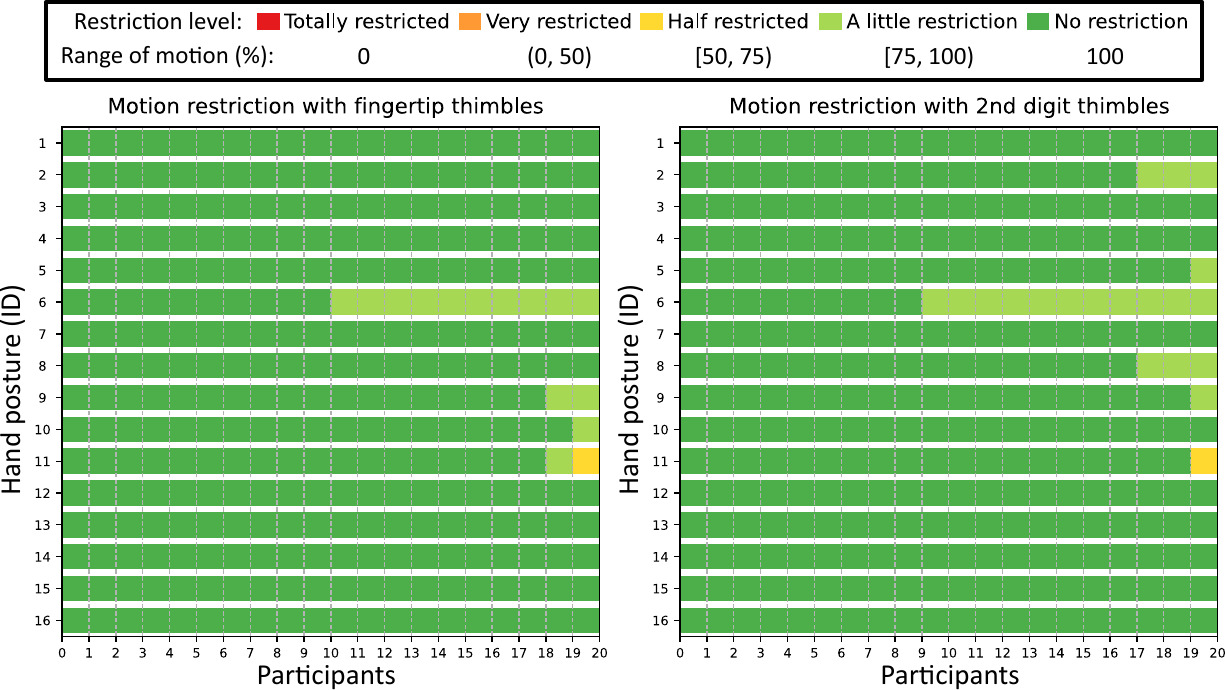}

    \vspace{0.3em}
    \subfiglab{fig:test1_handROM_results}\;
  \end{minipage}

  \caption{Range-of-motion (RoM) test: (a) Hand postures, and (b) Motion restrictions with each thimble attachment in comparison with the free hand.}
  \label{fig:RoM_overview_results}
\end{figure*}

This experiment evaluated the exoskeleton’s adaptability to different hand sizes and range of motion, as well as the effect of its structure (without feedback) on real-object manipulation. Two hypotheses were tested: (1) the exoskeleton adapts to a wide range of hand sizes while preserving natural digit motion, and (2) the strap thimble enables effective manipulation. For (1), objective metrics included 25 right-hand measurements (Fig.~\ref{fig:test1_visual_anatomy}, Table~\ref{tab:test1_hand_anatomy}), with a subset compared to ANSUR data \cite{greiner1991hand} to estimate population coverage. Digit motion was evaluated across 16 predefined hand poses under three conditions (free hand, exoskeleton with fingertip and strap thimbles), and assessed qualitatively through restriction levels assigned by the experimenter based on visual estimation of posture completion (Fig.~\ref{fig:RoM_overview_results}(\ref{fig:test1_handROM_results})). For (2), objective metrics included a customized Six-Hole Peg test adapted from the Nine-Hole Peg Test~\cite{Mathiowetz_1985}. Participants performed three repetitions with the free hand and with the exoskeleton attached at the second phalanx, after a short training period. Performance was quantified by completion time, defined as the time required to place all pegs and return them to their original positions, and peg-drop count, defined as pegs falling outside the target holes. Subjective evaluation included NASA-TLX \cite{hart2006nasa} for each condition in the peg test, and a customized set of 16 ergonomic questions (Table~\ref{tab:test1_questionnaire}).

\begin{figure}[t]
  \centering
  \includegraphics[width=\linewidth]{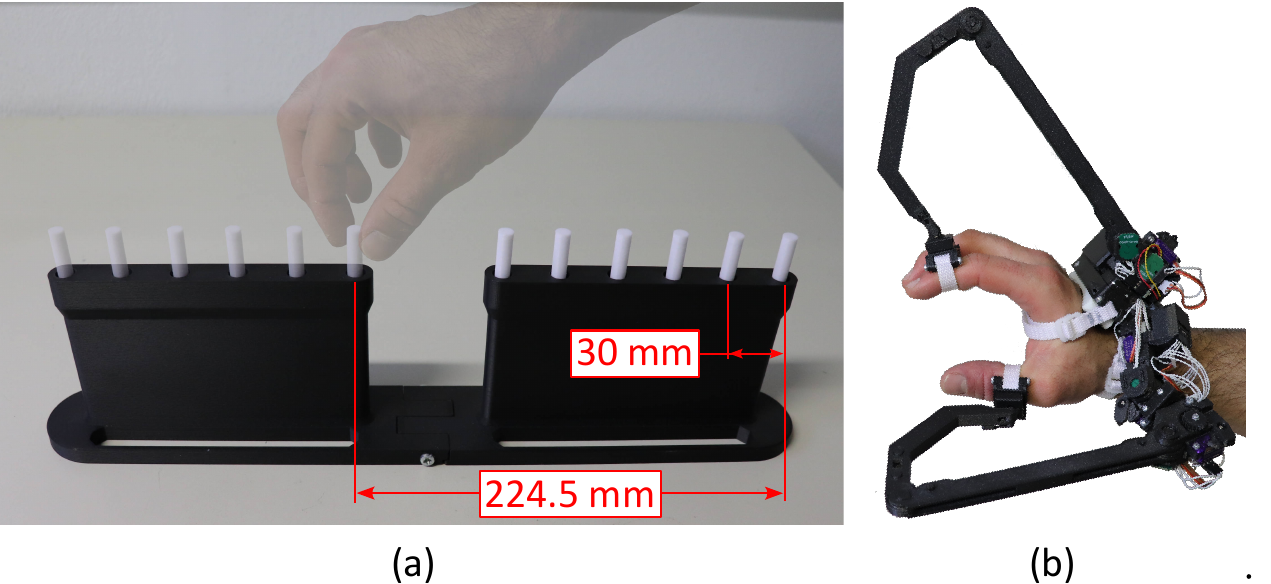}
  \caption{Setup for the Six-Hole Peg task: (a) Peg assembly, and (b) Exoskeleton with strap thimbles at the second phalanx.}
  \label{fig:PEGsetup}
\end{figure}

\begin{table}[t]
\caption{Ergonomics Questionnaire and Results (Study 1)}
\label{tab:test1_questionnaire}
\centering
\scriptsize
\setlength{\tabcolsep}{3pt}
\begin{tabular}{@{}r p{0.72\linewidth} r@{}}
\toprule
\textbf{ID} & \textbf{Question} & \textbf{Mdn [IQR]} \\
\midrule
1  & How well does the glove fit your hand? & 80 [73--91.5] \\
2  & How well does the tip thimble fit on your index finger? & 78 [67.5--90] \\
3  & How well does the tip thimble fit on your thumb? & 76.5 [63--89.25] \\
4  & How well does the strap thimble fit on the second digit of your index? & 85 [76.75--96] \\
5  & How well does the strap thimble fit on the second digit of your thumb? & 80 [65.75--93.25] \\
6  & How would you rate the weight of the exoskeleton? & 61 [48.25--67] \\
7  & To what extent did the exoskeleton restrict your motion in the index? & 10.5 [5.25--25.25] \\
8  & To what extent did the exoskeleton restrict your motion in the thumb? & 23 [13.75--41.75] \\
9  & To what extent did the exoskeleton restrict overall hand motion? & 23.5 [9--25.25] \\
10 & How stable did the exoskeleton feel while moving your fingertip thimbles? & 75 [48.75--89.25] \\
11 & How stable did the exoskeleton feel while moving your strap thimbles? & 82.5 [75--90.25] \\
12 & How comfortable was the exoskeleton material on your skin? & 67 [50--81] \\
13 & Did your hand feel hot or sweaty while wearing the exoskeleton? & 2 [0--28.25] \\
14 & To what extent did the exoskeleton affect performance in the Peg Test? & 38 [14--64.25] \\
15 & How would you compare your speed in the Peg Test with and without the exoskeleton? & 26 [20--35.25] \\
16 & How would you rate the overall comfort of the exoskeleton? & 70 [50--84.5] \\
\bottomrule
\end{tabular}
\vspace{2pt}
\footnotesize\par
\textit{Scales}: 0--100. Intermediate anchors (25, 75) were also provided.
Q1--5 fit (0=no fit, 50=acceptable, 100=perfect);
Q6 weight (0=very light, 50=neutral, 100=very heavy);
Q7--9 restriction (0=none, 50=some, 100=extreme);
Q10--11 stability (0=very unstable, 50=neutral, 100=fully stable);
Q12 \& Q16 comfort (0=very uncomfortable, 50=neutral, 100=very comfortable);
Q13--14 extent (0=none, 50=moderately, 100=extremely);
Q15 speed (0=much faster free hand, 50=same, 100=much faster exo).
\end{table}

\subsubsection{Experiment results}
The study included 20 participants (19 males, 1 female), aged 24 to 44 years (mean = 32.05). Seventeen participants were right-handed and three left-handed. Table~\ref{tab:test1_hand_anatomy} summarizes the hand measurements obtained in our study. To assess generalizability, these results were compared with anthropometric statistics from the ANSUR database \cite{greiner1991hand}. Because some of our measurements were not directly compatible due to different measurement definitions, only the overlapping variables were extracted and compared (Table~\ref{tab:test1_hand_comparison}). As the sample is predominantly male (95~\%), only the male data of both datasets were compared, while the inclusion of a single female participant is acknowledged as a limitation of the sample. As the ANSUR dataset reports percentiles for each parameter (p1–p99), inverse interpolation was used to estimate the percentiles corresponding to the observed minimum and maximum measurements, yielding population coverage (\%). Across the compared parameters, the minimum estimated coverage was 52.12\%, representing a lower bound and indicating that, for the most restrictive parameter, approximately half of the male ANSUR population is represented by our dataset.

\begin{table}[!htbp]
\caption{Comparison with Male ANSUR Data~\cite{greiner1991hand}}
\label{tab:test1_hand_comparison}
\centering
\scriptsize
\setlength{\tabcolsep}{4pt}
\begin{tabular}{@{}rccccccccc@{}}
\toprule
 & \multicolumn{2}{c}{Measured (male, mm)}
 & \multicolumn{4}{c}{Greiner\cite{greiner1991hand} (male, mm)}
 & \multicolumn{3}{c}{Comparison results (\%)} \\
\cmidrule(lr){2-3}\cmidrule(lr){4-7}\cmidrule(lr){8-10}
\textbf{ID} 
& Min & Max
& Mean & SD & Min & Max
& \textit{pMin} & \textit{pMax} & \% \\
\midrule
1  & 22.22 & 28.58 & 28.4 & 2.3 & 21 & 36 & $<1$ & 53.12 & 52.12* \\
2  & 20.98 & 32.53 & 22.6 & 2.4 & 16 & 32 & 24.98 & $>99$ & 74.02* \\
9  & 26.65 & 35.45 & 34.5 & 2.6 & 27 & 45 & $<1$ & 64.26 & 63.26* \\
12 & 79.43 & 91.15 & 90.4 & 4.2 & 79 & 106 & $<1$ & 57.09 & 56.09* \\
14 & 55.98 & 69.72 & 65.8 & 4.5 & 53 & 82 & 1.45 & 80.82 & 79.36 \\
16 & 51 & 60 & 57.4 & 1.6 & 53 & 65 & $<1$ & 94.79 & 93.79* \\
18 & 62 & 71 & 68.4 & 1.8 & 64 & 74 & $<1$ & 92.57 & 91.57* \\
21 & 62 & 73 & 72.3 & 2.9 & 63 & 81 & $<1$ & 59.54 & 58.54* \\
23 & 193 & 221 & 213.9 & 9.8 & 182 & 247 & 1.65 & 76.56 & 74.91 \\
25 & 159 & 188 & 174.3 & 8.2 & 143 & 204 & 3.1 & 95.26 & 92.16 \\ 
27 & 64.34 & 82.79 & 75.3 & 4.9 & 58 & 92 & 1.27 & 93.68 & 92.42 \\
\bottomrule
\end{tabular}
\vspace{3pt}
\footnotesize 
\par
\% $=$ \textit{pMax} $-$ \textit{pMin}. * Lower-bound coverage computed using p1 and p99 when values fall outside the ANSUR percentile range.
\end{table}

In the motion tests, restriction levels across the 16 postures varied with thimble type (Fig.~\ref{fig:RoM_overview_results}(\ref{fig:test1_handROM_results})). Participants were able to perform 12 and 10 out of 16 hand postures, respectively, without restriction. Limitations differed between thimble types and occurred predominantly near the end of the motion range. With the exception of one participant who reported moderate restriction in posture~11, all restrictions were minor. Slight restrictions were observed in posture~6 due to contact between the second linkage and the thimbles during full thumb radial abduction, and in a small number of cases in posture~8 due to interaction between the strap thimble and the palmar attachment strap. For postures~9–11, a slightly improved range of motion was observed with the strap thimble compared to the fingertip thimble, attributed to the latter being more likely to reach the limits of the exoskeleton motion. Overall, the exoskeleton allows most natural hand postures, with remaining limitations primarily associated with extreme configurations outside typical manipulation gestures (13–16), which were performed without restriction.

In the Six-Hole Peg test, no drops occurred with the free hand, and only four drops were observed with the exoskeleton (out of 60 rounds, 720 pick-and-place actions), each by a different participant in the third round. As drops were rare, no statistical analysis was conducted on drop counts. Execution times, however, differed significantly between conditions. For each subject and condition, times were averaged across three repetitions, and medians and IQRs were used due to non-normality. Median task time was $19.99$~s (IQR $= 15.69$–$22.78$) with the exoskeleton and $14.60$~s (IQR $= 12.79$–$16.21$) with the free hand, yielding a median paired difference of $4.70$~s, 95\% CI [$3.52$, $6.80$]. A one-sided Wilcoxon signed-rank test confirmed longer times with the exoskeleton, $W = 210$, $z = 3.91$, $p < .001$, with a large effect size ($r = .88$). For context, mean times were $14.89 \pm 2.64$~s (free hand) and $20.48 \pm 4.94$~s (exoskeleton), corresponding to a mean increase of $5.58 \pm 2.71$~s, or $36.5 \pm 13.5$\%. During this test, the exoskeleton was used passively, without applying any force feedback. Under this condition, longer execution times were observed compared to the free hand condition. The increased execution times may reflect constraints introduced by the exoskeleton’s structure and kinematics. Nonetheless, aside from the four isolated drops, participants were still able to complete the manipulation task with the exoskeleton.

\begin{figure}[htbp]
  \centering
  \includegraphics[width=\linewidth]{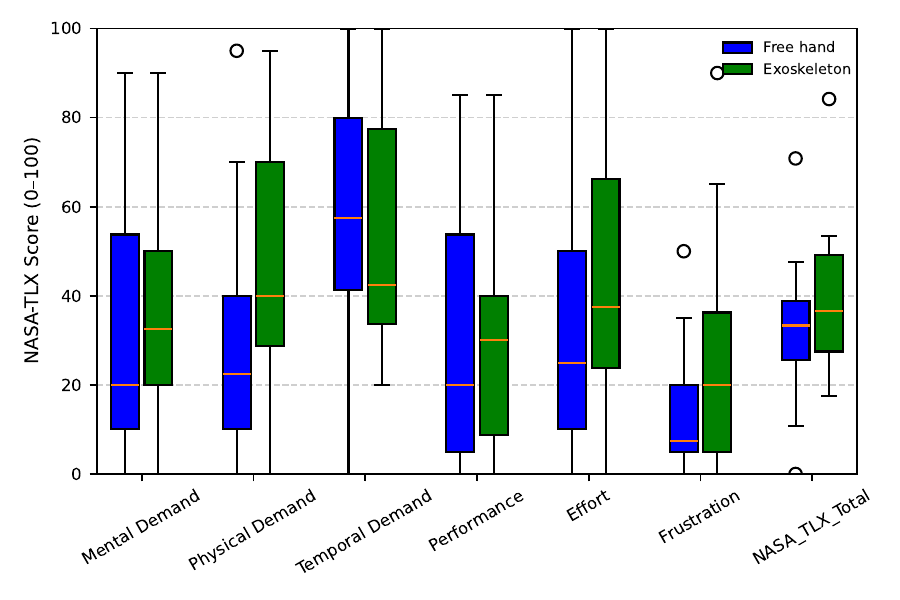}
  \caption{NASA-TLX results for the Six-Hole Peg task.}
  \label{fig:test1_NASATLX_Plot}
\end{figure}

NASA-TLX scores were compared between free hand and exoskeleton conditions using one-sided Wilcoxon signed-rank tests, under the alternative hypothesis that workload would be higher with the exoskeleton (Fig.~\ref{fig:test1_NASATLX_Plot} and Table~\ref{tab:test1_NASATLX_results}). After Bonferroni-Holm correction, the exoskeleton condition was rated significantly higher in mental demand, physical demand, effort, frustration and overall workload (all $p < .05$, \text{one-sided}). Temporal demand and performance showed no significant differences (p = .89 and .5, respectively), despite objectively measured metrics indicating otherwise.

\begin{table}[htbp]
\caption{NASA-TLX scores for the Six-Hole Peg Test (median [IQR]) and Wilcoxon test results (n=20).}
\label{tab:test1_NASATLX_results}
\centering
\scriptsize
\setlength{\tabcolsep}{2pt} 
\begin{tabular}{@{}l c c p{10.5em}@{}}
\toprule
\textbf{Subscale} & \textbf{Free} & \textbf{Exo} & \textbf{Wilcoxon (W, z, p, r)} \\
\midrule
Mental Demand     & 20 [10.0-53.75] & 32.5 [20.0-50.0] & 113.0, 2.39, .0410, .59 \\
Physical Demand   & 22.5 [10.0-40.0] & 40.0 [28.75-70.0] & 166.0, 3.57, .0016, .84 \\
Temporal Demand   & 57.5 [41.25-80.0] & 42.5 [33.75-77.5] & 33.0, -1.23, .8966, .33 \\
Performance       & 20.0 [5.0-53.75] & 30.0 [8.75-40.0] & 63.5, 0.67, .5081, .18 \\
Effort            & 25.0 [10.0-50.0] & 37.5 [23.75-66.25] & 135.5, 2.20, .0458, .51 \\
Frustration       & 7.5 [5.0-20.0] & 20.0 [5.0-36.25] & 84.0, 2.71, .0193, .75 \\
NASA-TLX Total    & 33.33 [25.625-38.75] & 36.66 [27.5-49.16] & 168.0, 2.93, .0105, .67 \\
\bottomrule
\end{tabular}
\end{table}

The ergonomics questionnaire results (Table~\ref{tab:test1_questionnaire}) can be grouped into several categories. \textbf{Fit} (Q1--5) was rated positively, with medians between 76.5 and 85, indicating generally good adaptation of the glove and thimbles. \textbf{Weight} (Q6) was judged moderately heavy (Mdn = 61, IQR = 48.25--67), with most values between neutral and heavy. \textbf{Restriction} (Q7--9) was low overall: the index finger showed the least limitation (Mdn = 10.5), followed by the thumb (Mdn = 23), while overall hand restriction remained below 50 for almost all participants. \textbf{Stability} (Q10--11) was consistently high, with both fingertip and strap thimbles rated mostly stable (Mdn $\geq$ 75). \textbf{Comfort} (Q12, Q16) was also positive (Mdn = 67--70), although not maximal. \textbf{Heat and sweating} (Q13) were reported as minimal (Mdn = 2, IQR = 0--28.25). Finally, participants acknowledged an impact on perceived performance in the Six-Hole Peg test, rating completion as faster with the free hand (Q14--15, Mdn = 26--38), consistent with the observed execution times.

Overall, this first test evaluated ergonomics, mobility and the effects of wearing the exoskeleton. The device enabled most movements with minimal restriction in a group of subjects that, for the comparable parameters, represented at least approximately half of the ANSUR male population based on the most restrictive anthropometric measure, while offering satisfactory yet improvable ergonomics. With regard to movement and task performance, the Six-Hole Peg task showed increased execution time when using the exoskeleton, but very low impact on effectiveness (low failure rate).

\subsection{Pick-and-place manipulation task}
\subsubsection{Experiment definition}
\begin{figure}[htbp]
  \centering
  \includegraphics[width=0.9\linewidth]{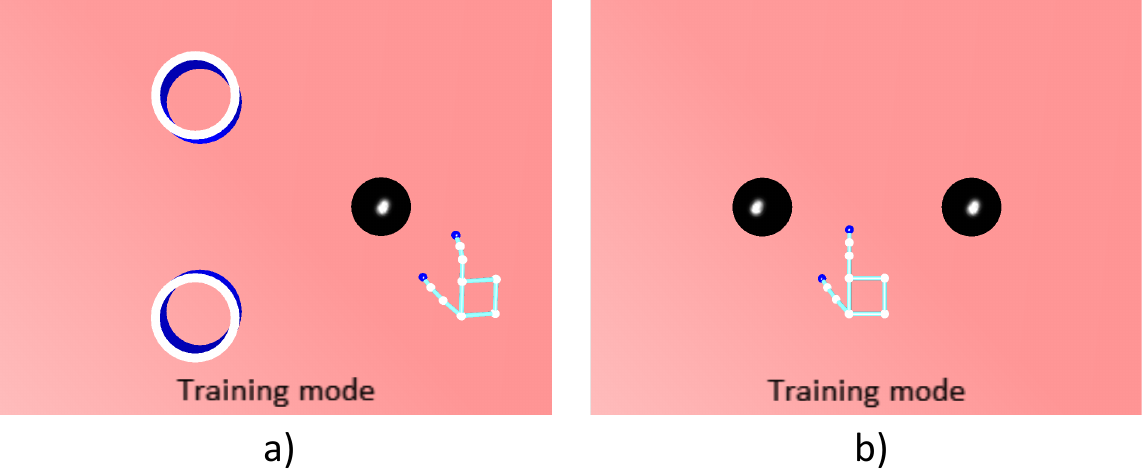}
  \caption{Virtual setups for (a) the pick an place test and (b) perceived softness discrimination experiment.}
  \label{fig:tests2and3_setup}
\end{figure}

The second experiment assessed the influence of force feedback from the exoskeleton’s ratchet–pawl mechanism compared to free hand control in a virtual pick-and-place task. Participants were required to grasp a virtual sphere and place it into one of two target rings (Fig.~\ref{fig:tests2and3_setup}(a)). The task was implemented in CHAI3D, which issued binary motion-blocking commands to the glove upon virtual contact detection (on: feedback engaged, off: free motion). The independent variables were:
\begin{itemize}
\item \textbf{Hand condition:} free hand tracked with the Leap Motion controller, versus exoskeleton, where the Leap Motion provided global hand tracking while the exoskeleton supplied finger sensing and force feedback.
\item \textbf{Tolerance threshold for failure:} two values (0.025 and 0.03). During object pickup, fingertip distance was recorded and the threshold defined the allowable variation. Smaller thresholds increased the likelihood of dropping or over-squeezing the object (thus higher difficulty). These values were set after pilot testing and are not meant to represent real-world dimensions.
\item \textbf{Target ring:} far (Goal 1) versus near (Goal 2) from the subject, both aligned along the forward (Y) axis and equidistant from the initial ball position on the right.
\item \textbf{Repetitions:} Three per stimulus, resulting in 12 samples per hand condition and 24 in total. 
\end{itemize}

The hypothesis was that force feedback from the exoskeleton would reduce execution time and failure rates and yield more favorable subjective ratings, compared to free hand interaction in virtual manipulation. Objective metrics included execution time per trial, defined as the time from initial object contact to successful placement, and failures, defined as dropping or exceeding the tolerance threshold. Subjective metrics consisted of NASA-TLX for each hand condition and a questionnaire on force feedback, acoustic perception, and overall preference (Table~\ref{tab:test2_questionnaire}).

\begin{table}[htbp]
\caption{Questionnaire (Study 2)}
\label{tab:test2_questionnaire}
\centering
\begin{tabular}{@{}cp{0.9\linewidth}@{}}
\toprule
ID & Question \\ \midrule
1  & To what extent was the force feedback helpful during the task? \\
2  & How noticeable were the sounds coming from the exoskeleton actuation system? \\
3  & How distracting were the motor sounds? \\
4 & How would you rate the annoyance of the motor sounds? \\
5 & To what extent were the sounds from the exoskeleton motors helpful during the task? \\
6 & Which condition do you prefer for this type of task: with the free hand or with the exoskeleton? \\
\bottomrule
\end{tabular}
\vspace{4pt}
\footnotesize 
\par
\textit{Scales}: 0--100. Intermediate anchors (25, 75) were also provided.\\
Q1, Q5: 0 = Not, 50 = Moderate, 100 = Extremely. \\
Q2: 0 = Not, 50 = Somewhat, 100 = Very noticeable. \\
Q3, Q4: 0 = None, 50 = Moderate, 100 = Extreme. \\
Q6: 0 = Strong free hand, 50 = No preference, 100 = Strong exo \\
\end{table}

\subsubsection{Experiment results}
The study included 20 participants (16 males, 4 females), aged 26–62 years (mean = 32.25), of whom 18 were right-handed and two left-handed. For completion times, mean values were computed per participant for each combination of tolerance and goal. As data were not normally distributed, Wilcoxon signed-rank tests were applied under the alternative hypothesis that the free hand condition would yield longer times. Bonferroni–Holm correction was applied across the four comparisons. As summarized in Table~\ref{tab:test2_times}, all differences remained statistically significant after correction ($p$ range: $.0002–.0068$). The free hand condition consistently resulted in longer completion times ($\tilde{\Delta} \approx$ +0.4–0.6~s), with large effect sizes in all cases.

Across all trials, more failures occurred with the free hand (320) than with the exoskeleton (29). As failures per configuration (goal and threshold) were sparse and commonly zero for the exoskeleton, they were aggregated per participant across all trials. The free hand condition produced substantially more failures (Mdn = 9.5, IQR = 2.0–31.5) than the exoskeleton (Mdn = 1.0, IQR = 0.0–2.25). A one-sided Wilcoxon signed-rank test confirmed this difference, $W = 145$, $z = 3.29$, $p < .001$, with a large effect size ($r = .79$).

When comparing data between the two tolerance thresholds (0.025 vs. 0.03), no consistent differences were found. Occasional effects were observed in failures, but these were based on sparse and zero-inflated data and are not considered conclusive. Similarly, no statistically significant differences were observed between Goal 1 and Goal 2 after applying Wilcoxon tests with Bonferroni--Holm correction.

\begin{table}[htbp]
\caption{Wilcoxon results for completion times (Leap Motion vs Exoskeleton)}
\label{tab:test2_times}
\centering
\scriptsize
\setlength{\tabcolsep}{2pt}
\begin{tabular}{@{}l c c c c c c c@{}}
\toprule
Cfg & Exo [Mdn, IQR] & Leap [Mdn, IQR] & $\tilde{\Delta}$ [CI] & W & z & p & r \\
\midrule
G1--.03  & 1188 [1018--1577] & 1742 [1157--2344] & +540 [--6; 959]   & 179 & 2.75 & .0042 & .62 \\
G1--.025 & 1133 [1079--1400] & 1821 [1377--2635] & +611 [262; 977]   & 199 & 3.50 & .0002 & .78 \\
G2--.03  & 1144 [1058--1534] & 1885 [1153--2479] & +422 [--98;1174]  & 170 & 2.42 & .0068 & .54 \\
G2--.025 & 1158 [1041--1446] & 1982 [1162--2600] & +515 [66;1218]    & 182 & 2.87 & .0040 & .64 \\
\bottomrule
\end{tabular}
\vspace{2pt}
\footnotesize
Values are medians [IQR] in ms. $\tilde{\Delta}$ = median paired difference (Exo--Leap). CI = 95\%. 
Cfg = configuration: G1/G2 = target goal, .025/.03 = tolerance threshold. 
$p$ values are Bonferroni--Holm corrected.
\end{table}

\begin{figure}[!t]
  \centering
  \includegraphics[width=1\linewidth]{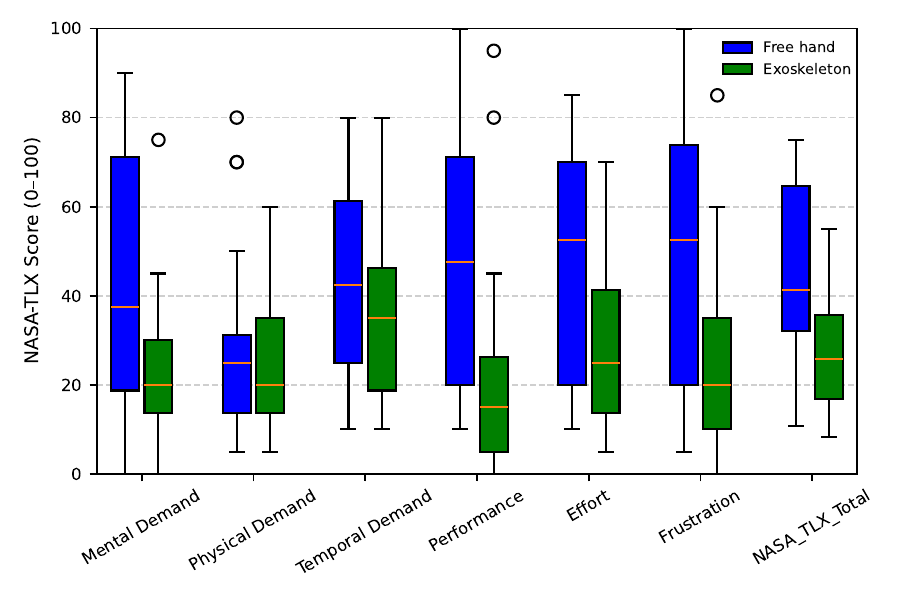}
  \caption{NASA-TLX results for the virtual pick-and-place task.}
  \label{fig:test2_NASATLX_Plot}
\end{figure}

NASA-TLX scores were compared between free hand and exoskeleton conditions using one-sided Wilcoxon signed-rank tests, under the alternative hypothesis that workload would be higher with the free hand. Results are summarized in Fig.~\ref{fig:test2_NASATLX_Plot} and Table~\ref{tab:test2_NASATLX_results}. After Bonferroni-Holm correction, significant differences were observed in Mental Demand ($p = .008$), Performance ($p = .02$), Effort ($p = .01$), Frustration ($p = .02$) and the overall NASA-TLX Total score ($p = .01$), all favoring the exoskeleton. Physical Demand and Temporal Demand did not reach significance ($p = .155$ each). These results suggest that the exoskeleton condition was perceived as less demanding overall, with lower mental demand, frustration, and effort ratings, as well as better perceived performance.

\begin{table}[!t]
\caption{NASA-TLX scores of virtual pick-and-place test reported as median [IQR] and Wilcoxon signed-rank results (n=20).}
\label{tab:test2_NASATLX_results}
\centering
\scriptsize
\setlength{\tabcolsep}{2pt} 
\begin{tabular}{@{}l c c p{10.5em}@{}}
\toprule
\textbf{Subscale} & \textbf{Free} & \textbf{Exo} & \textbf{Wilcoxon (W, z, p, r)} \\
\midrule
Mental Demand     & 37.5 [18.75--71.25] & 20.0 [13.75--30.0] & 114.0, 3.08, .008, .80 \\
Physical Demand   & 25.0 [13.75--31.25] & 20.0 [13.75--35.0] & 114.5, 1.27, .155, .30 \\
Temporal Demand   & 42.5 [25.0--61.25]  & 35.0 [18.75--46.25] & 118.5, 1.46, .155, .34 \\
Performance       & 47.5 [20.0--71.25]  & 15.0 [5.0--26.25]   & 172.5, 2.53, .020, .57 \\
Effort            & 52.5 [20.0--70.0]   & 25.0 [13.75--41.25] & 153.5, 2.98, .010, .70 \\
Frustration       & 52.5 [20.0--73.75]  & 20.0 [10.0--35.0]   & 159.5, 2.60, .020, .60 \\
NASA-TLX Total    & 41.25 [32.08--64.58] & 25.83 [16.88--35.63] & 153.0, 2.93, .010, .69 \\
\bottomrule
\end{tabular}
\end{table}

In the custom questionnaire, participants rated force feedback as helpful (Q1: Mdn = 64.5, IQR = 40.0–75.0). Sounds were clearly noticeable (Q2: Mdn = 75.0, IQR = 25.8–76.3), but rated low in distraction (Q3: Mdn = 5.5, IQR = 1.5–23.5) and annoyance (Q4: Mdn = 5.5, IQR = 2.8–12.8). The perceived helpfulness of sounds varied considerably across participants (Q5: Mdn = 27.5, IQR = 7.8–70.3). Finally, overall preference strongly favored the exoskeleton (Q6: Mdn = 85.0, IQR = 60.8–97.8), well above the neutral midpoint of 50.

These results indicate that the force feedback condition improved task performance compared to the free hand condition, yielding shorter completion times and fewer failures. Subjective measures aligned with these findings: workload (NASA-TLX) was generally lower, force feedback was perceived as helpful, and motor sounds were noticeable but not distracting. Overall user preference favored the exoskeleton, supporting its effectiveness and acceptance in virtual pick-and-place tasks.

\subsection{Perceived softness discrimination study}
\subsubsection{Experiment definition}
This experiment evaluated the one-way clutch mechanism in a pinching softness discrimination task in VR. Participants were instructed to probe two virtual spheres per stimulus and indicate which one felt softer (Fig.~\ref{fig:tests2and3_setup}(b)). In this study, softness refers to the perceived compliance of the virtual object, arising from different position-based motion resistance profiles rendered by the exoskeleton, rather than from physically calibrated stiffness values. The fingertip position in CHAI3D was updated using AS5600 angle measurements from the physical exoskeleton. Upon contact, CHAI3D sent an activation signal and the fingertip penetration depth to the ESP32. This value was used in a closed-loop position control algorithm running on the ESP32 to generate unidirectional active resistance on the digits. The position command for each servomotor is computed as

\begin{equation}
\theta_{i,\mathrm{cmd}} \;=\; \theta_{\mathrm{eq}} \;+\; 0.01\,c\,s_i,
\label{eq:softness_controller}
\end{equation}
where $\theta_{i,\mathrm{cmd}}$ is the servomotor position command expressed in internal control units (0–600~uds), corresponding to a physical motion range of 0 to $120^\circ$. The variable \(s_i\) is the virtual fingertip indentation into the object expressed in simulation units defined by the virtual object geometry and hand model (1: surface contact, 295: full pinch closure), \(c\) is an empirically calibrated compliance parameter, \(0.01\) is a fixed gain, and \(\theta_{\mathrm{eq}}\) is the equilibrium motor position when the clutch completes engagement and feedback is activated. Three compliance levels were tested, selected via preliminary calibration: rigid (A, $c=10$), medium-rigid (B, $c=50$) and soft (C, $c=90$). Lower values of \(c\) restrict indentation and emulate higher stiffness, whereas higher values of \(c\) allow deeper indentation. 

The hypothesis was that the system would render perceptually distinct softness levels, enabling discrimination above chance (\>50\%). All pairwise comparisons (AB, AC, BC) were tested with three repetitions each, yielding nine samples per participant, with randomized stimulus order and left/right placement. Prior to the task, participants performed a brief training (2~min) to become familiar with the procedure. Data collection included both objective and subjective measures. Objective data consisted of discrimination accuracy, defined as the proportion of correct softer-ball selections per condition. Subjective data included a short questionnaire on task perception, force feedback and acoustic awareness (Table~\ref{tab:test3_questionnaire}).

\begin{table}[htbp]
\caption{Questionnaire (Study 3)}
\label{tab:test3_questionnaire}
\centering
\setlength{\tabcolsep}{2pt}
\begin{tabular}{@{}cp{0.95\linewidth}@{}}
\toprule
ID & Question \\ \midrule
1 & To what extent could you perceive differences in stiffness between the two balls in each trial? \\
2 & How confident were you in identifying the softer ball in each comparison? \\
3 & How often did the two balls feel indistinguishable in terms of stiffness? \\
4 & How noticeable were the sounds coming from the exoskeleton actuation system? \\
5 & To what extent did the motor sounds distract you from the task? \\
6 & How would you rate the annoyance of the motor sounds? \\
7 & To what extent were the sounds from the exoskeleton motors helpful during the task? \\
\bottomrule
\end{tabular}
\vspace{4pt}
\footnotesize 
\par
\textit{Scales}: 0--100. Intermediate anchors (25, 75) were also provided. \\
Q1: 0 = No difference, 50 = Moderate differences, 100 = Extremely clear. \\
Q2: 0 = Not confident, 50 = Moderate, 100 = Completely. \\
Q3: 0 = Never, 50 = Sometimes, 100 = Always. \\
Q4: 0 = Not noticeable, 50 = Somewhat, 100 = Very. \\
Q5-7: 0 = Not at all, 50 = Moderate, 100 = Extremely. \\

\end{table}

\subsubsection{Experiment results}
The study included 20 participants (18 males, 2 females), aged 24–39 years (mean = 30.15), of whom 14 were right-handed and 6 left-handed. Performance was assessed as correct or incorrect identification of the softer object. Across all trials, the overall discrimination accuracy was high (Mdn = 88.9\%, IQR = 77.8–100.0). As the data were not normally distributed, nonparametric analyses were applied. Because responses were binary (success vs. failure), one-sided binomial tests were conducted against the null hypothesis of chance performance (50\%), with Bonferroni-Holm correction across the three pairwise comparisons (AB, AC, BC). Results are summarized in Table~\ref{tab:test3_performance_results}. Accuracy was statistically significant above chance for all comparisons (all $p < .001$), with large to very large effect sizes (Cohen’s $h = 0.69$–$0.99$), indicating that participants reliably discriminated between the tested softness levels. 

\begin{table}[htbp]
\caption{Binomial test results for softness comparisons}
\label{tab:test3_performance_results}
\centering
\scriptsize
\setlength{\tabcolsep}{4pt}
\begin{tabular}{@{}l c c c c c@{}}
\toprule
Pair & Correct & Proportion & Clopper–Pearson 95\% CI & $p$ (Holm) & Cohen’s $h$ \\
\midrule
AB & 49/60 & 0.82 & [0.71, 1.00] & $<.001$ & 0.69 (large) \\
AC & 55/60 & 0.92 & [0.83, 1.00] & $<.001$ & 0.99 (very large) \\
BC & 50/60 & 0.83 & [0.73, 1.00] & $<.001$ & 0.73 (large) \\
\bottomrule
\end{tabular}
\end{table}

In the custom questionnaire, participants generally reported that stiffness differences between the two balls were clearly perceptible (Q1: Mdn = 67, IQR = 48.3–75.0) and expressed moderate to high confidence in identifying the softer ball (Q2: Mdn = 70, IQR = 47.5–85.0). Still, indistinguishability was occasionally reported (Q3: Mdn = 33.5, IQR = 24.8–54.8), indicating that some comparisons remained challenging. Motor sounds were rated as highly noticeable (Q4: Mdn = 75.0, IQR = 58.3–90.3), but distraction (Q5: Mdn = 10.0, IQR = 0.0–18.5) and annoyance (Q6: Mdn = 14.0, IQR = 0.0–23.5) were minimal. Finally, the perceived helpfulness of sounds varied substantially (Q7: Mdn = 25.0, IQR = 9.3–52.0), suggesting that while audible cues were sometimes useful, their contribution was not consistent across participants.

Overall, results indicate that the one-way clutch mechanism rendered perceptually distinct compliance levels that participants reliably discriminated in pairwise comparisons. Subjective ratings indicated moderately high perception of stiffness differences and confidence in judgments, while acoustic feedback was noticeable but not distracting or annoying. Taken together, these findings support the hypothesis that the proposed system can provide perceptually meaningful unidirectional force feedback for softness discrimination. As no objective interaction measures were recorded, conclusions are limited to subjective perception and relative discriminability.

\section{Discussion}
The experimental evaluations present positive results regarding the intended design goals and align with prior haptic glove studies. Ergonomics, a key focus in commercial systems such as SenseGlove and HaptX, were evaluated favorably, demonstrating adaptation to different hand sizes and largely unrestricted motion across tested gestures at both fingertip and second-phalanx attachment points. However, larger and more gender-balanced studies are required to assess broader applicability. User studies confirmed perceptually meaningful force feedback for both actuation mechanisms. The ratchet–pawl mechanism provides rigid motion constraints similar in principle to those in prior systems such as Dexmo and Wolverine, although stiffness was not quantitatively evaluated, while the third study showed that the clutch mechanism can render perceptually distinct compliance levels. In the pick-and-place task, failure rates were lower with force feedback compared to the free-hand condition, indicating improved grasp control. This is consistent with prior work on Dexmo~\cite{gu2016dexmo} in VR and Leonardis et al.~\cite{leonardis_hand_2024} in teleoperation, where task success or error rates improved with force feedback. The latter also reported reduced execution times despite differences in the experimental setup.

At the system level, commercial devices such as HaptX G1 and SenseGlove Nova 2 have comparable weight to KinesCeTI (450 g and 350 g vs. 376 g) and lower latency (HaptX G1: 62 ms vs. 83.15 ms ratchet–pawl, 100.1 ms clutch) but offer limited force feedback dimensionality. Linkage-based systems in teleoperation and rehabilitation \cite{leonardis_hand_2024,bartalucci_original_2023,iqbal_four-fingered_2015} provide effective force feedback but rely on external, immovable actuation, limiting portability, whereas KinesCeTI supports multiple feedback modalities in a portable, wearable architecture.

Adaptability to different users has been addressed in prior work through adjustable components \cite{chinello_modular_2020}, multiple glove sizes (HaptX, Senseglove), or size-adaptable exoskeleton designs \cite{iqbal_four-fingered_2015}. The proposed linkage structure extends this adaptability by supporting attachment at both the fingertip and the second phalanx, whereas prior designs focus on a single contact location. Together, this dual-point adaptability and the consideration of anthropometric extremes introduce a key trade-off: increased volume. This increases the likelihood of collisions in constrained environments and contributes to mass and inertia that are perceptible to users, particularly due to extensive use of bearings and prototyped electronics. These effects could be mitigated through structural optimization and compact PCB-based electronics, which would also free dorsal space and enable extension to additional digits. The thumb assembly, despite incorporating a compression spring to reduce inertial effects, would particularly benefit from such refinements. As a grounded exoskeleton, the system transmits reaction forces to the hand, which remained noticeable despite a compliant TPU–neoprene attachment and could be further reduced through optimizations. Another limitation is the absence of integrated force sensing, necessary for closed-loop force control. The main pulley offers a natural location for integrating force sensing within future actuation modules.

Despite these limitations, KinesCeTI adopts a different approach from many existing solutions that rely on closed architectures with fixed capabilities or prioritize specific features at the expense of adaptability and wearability. Its contribution therefore lies not in outperforming existing systems in isolated metrics, but in providing a versatile and modular platform that enables researchers to integrate and evaluate novel haptic actuation and feedback concepts, thereby lowering the entry barrier for haptic research. Further contributions of this work include the actuation mechanisms, the explicit consideration of acoustic perception, and the evaluation protocols, all of which may be reused or adapted for other hand-based haptic interfaces, potentially extending the impact of this work. 

\section{Conclusions}
This work presents KinesCeTI, a modular and portable force feedback glove designed as a flexible research platform. It integrates an adaptive linkage structure, bidirectional tendon routing, and interchangeable thimble and actuation modules, enabling the exploration of multiple force feedback modalities. Two mechanisms were developed: a ratchet–pawl brake and a novel one-way clutch, first presented in this work. 

The system was evaluated in three user studies addressing ergonomics, hand motion, and interaction performance in real and virtual environments. The results demonstrated that KinesCeTI can be worn by users with different hand sizes while being attached at either the fingertip or the second phalanx, and that both actuation modules rendered perceptually meaningful force feedback during manipulation tasks.

Rather than targeting peak performance in individual metrics, KinesCeTI is conceived as an open and adaptable platform for haptics and VR research. Its modular architecture enables comparative testing of multiple actuation concepts within a single glove, either at the Knuckle Hubs or through interchangeable thimbles that allow integration of tactile, thermal, or multimodal feedback modules without structural redesign. This reduces the effort required to develop and systematically evaluate diverse haptic systems.

\section*{Acknowledgment}
Funded by the German Research Foundation (DFG, Deutsche Forschungsgemeinschaft) as part of Germany’s Excellence Strategy – EXC 2050/1 – Project ID 390696704 – Cluster of Excellence “Centre for Tactile Internet with Human-in-the-Loop” (CeTI) of Technische Universität Dresden.

\bibliographystyle{IEEEtran}
\bibliography{KinesCeTIReferences2025_vfixed}
\begin{IEEEbiography}[{\includegraphics[width=1in,height=1.25in,clip,keepaspectratio]{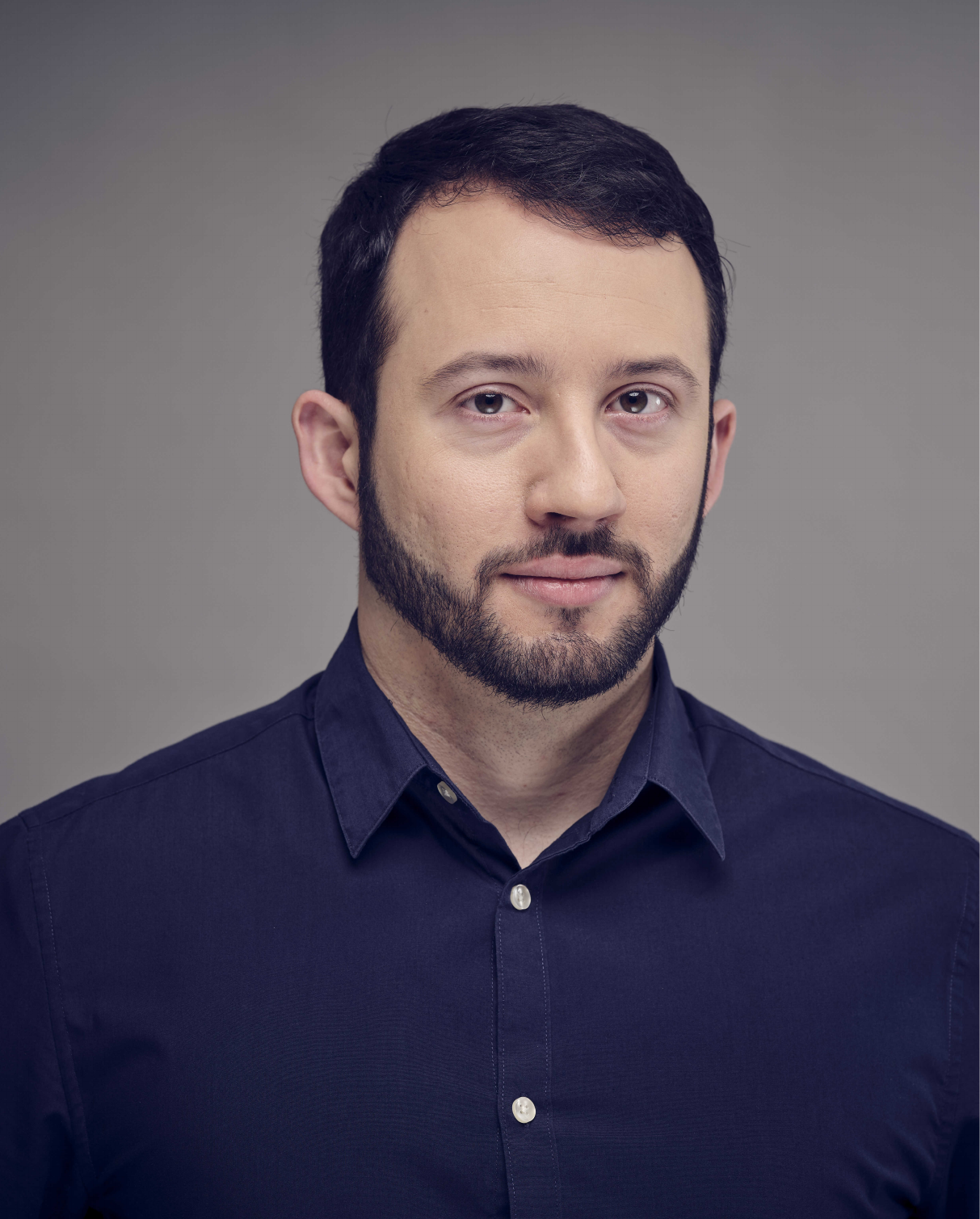}}]{Pablo Alvarez Romeo} received the M.Sc. degree in Electronic Engineering  from the University of Zaragoza, Zaragoza, Spain. He is currently a Doctoral researcher with the Chair of Acoustics and Haptics, TU Dresden. His research interests include haptic interfaces and perception, robotics, mechatronics design and development, wearables, additive manufacturing and rapid prototyping.
\end{IEEEbiography}
\begin{IEEEbiography}[{\includegraphics[width=1in,height=1.25in,clip,keepaspectratio]{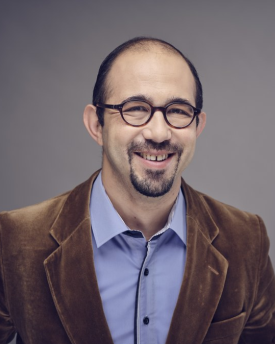}}]{Mehmet Ercan Altinsoy} received his degree in mechanical engineering from the Technical University of Istanbul and his Ph.D. in electrical engineering from Ruhr-University Bochum in 2005. After working as a consulting engineer at HEAD Acoustics, he joined TU Dresden in 2006, where he is currently Professor in Acoustic and Haptic Engineering. His research focuses on perception-based engineering, vibroacoustics, vehicle acoustics, and haptic interfaces. In 2018, he was awarded a Visiting Professorship from Tohoku University, Japan. He is a Lothar-Cremer Medalist of DEGA, Chairman of the DEGA Vehicle NVH Expert Committee, and a core member of the Cluster of Excellence Centre for Tactile Internet with Human-in-the-Loop.
\end{IEEEbiography}

\end{document}